\documentclass[aps,prc,amsfonts,preprintnumbers,superscriptaddress,nofootinbib]{revtex4}
\usepackage{wick}
\usepackage{graphics}
\usepackage{amsmath}
\usepackage{hyperref}
\usepackage{amssymb}
\usepackage{psfrag}
\usepackage{epsfig}
\usepackage{epsf}
\usepackage{float}
\usepackage{slashed}
\usepackage{dsfont}
\usepackage{bbold}
\usepackage{color}
\allowdisplaybreaks

\newcommand{\fet}[1]{\mbox{\boldmath $#1$}}

\newcommand{\beq}{\begin{equation}}
\newcommand{\eeq}{\end{equation}}
\newcommand{\beqa}{\begin{eqnarray}}
\newcommand{\eeqa}{\end{eqnarray}}
\newcommand{\nn}{\nonumber \\ }

\def\sumint{\sum\hspace{-14.0pt}\int_{\hspace{-11.0pt} \begin{array}{c} \\[-3pt] \scriptstyle nrs\end{array}} \hspace{0.0pt}}
\def\sumintnum{\sum\hspace{-14.0pt}\int_{\hspace{-10.0pt} \begin{array}{c} \\[-1pt] \scriptstyle 23\end{array}} \hspace{5.0pt}}

\begin{document}

\title{Towards consistent nuclear interactions from chiral
  Lagrangians I: \\The path-integral approach}

\author{H.~Krebs}
\email[]{Email: hermann.krebs@rub.de}
\affiliation{Institut f\"ur Theoretische Physik II, Ruhr-Universit\"at Bochum,
  D-44780 Bochum, Germany}
\author{E.~Epelbaum}
\email[]{Email: evgeny.epelbaum@rub.de}
\affiliation{Institut f\"ur Theoretische Physik II, Ruhr-Universit\"at Bochum,
  D-44780 Bochum, Germany}
\date{\today}

\begin{abstract}
Low-energy nuclear interactions have been extensively studied in the
framework of chiral effective field theory. The corresponding
potentials have been worked out
using dimensional regularization to
evaluate ultraviolet divergent loop integrals. An additional cutoff
is then introduced in the nuclear Schr\"odinger equation to
calculate observables. Recently, we have shown that
such a mixture of 
two regularization schemes
violates
chiral symmetry when applied beyond the two-nucleon system
and/or to processes involving external probes.
To solve this issue, three- and four-nucleon
forces as well as exchange current operators need to be re-derived
using symmetry-preserving cutoff regularization. While it is possible
to introduce a symmetry-preserving
cutoff
already in the effective chiral Lagrangian, the
appearance of high-order time derivatives of the pion field, caused by the 
regulator, makes the standard Hamiltonian-based methods not well suited
for the calculation of nuclear potentials. Here,
we propose a new approach to derive nuclear interactions using the
path integral method
with no reliance on the canonical quantization.
To this aim, the interaction part of the action is brought to an 
instantaneous form via suitably chosen nonlocal field
redefinitions. Loop contributions to the nuclear potentials are then
generated through the functional determinant, induced by the field
redefinitions. We discuss in detail the application of these ideas to
the case of 
a regularized Yukawa-type model of pion-nucleon interactions. Our new method allows to perform a
systematic quantum mechanical reduction 
within the quantum field theory framework and opens the way for deriving consistently
regularized nuclear forces and current operators
from the effective chiral Lagrangian.
\end{abstract}


\maketitle

\vspace{-0.2cm}

\section{Introduction}
\def\theequation{\arabic{section}.\arabic{equation}}
\label{sec:Intro}

Chiral effective field theory (EFT) is nowadays the most commonly used
approach to analyze low-energy nuclear structure and reactions.
It goes back to the seminal papers by Weinberg in the early
1990s \cite{Weinberg:1990rz,Weinberg:1991um}, which laid out the
foundations and set the agenda for the upcoming work along these
lines~\cite{Epelbaum:2008ga,Machleidt:2011zz}. In short terms, the
main idea is to apply chiral perturbation theory (ChPT) \cite{Bernard:1995dp,Bernard:2007zu,Scherer:2012xha} to calculate
the few-nucleon-irreducible part of the scattering amplitude, i.e.~that
part which 
cannot be generated from iterations of the Lippmann-Schwinger
equation. The set of all possible irreducible diagrams is identified
with the
nuclear Hamiltonian and current operators, which can be used to calculate
nuclear observables by solving the $A$-body Schr\"odinger equation. 

Derivation of nuclear interactions from the effective chiral
Lagrangian requires separating out the irreducible parts of the
amplitude and thus goes beyond the standard Feynman diagram calculus.
In the pioneering work by Weinberg
\cite{Weinberg:1990rz,Weinberg:1991um} and by
Ord\'{o}\~{n}ez et al.~\cite{Ordonez:1995rz}, 
old-fashioned time-ordered perturbation theory (TOPT) was employed,
see
Refs.~\cite{Pastore:2008ui,Pastore:2009is,Pastore:2011ip,Baroni:2015uza,deVries:2020iea,Baru:2019ndr}
for more recent applications of TOPT to derive nuclear potentials in
chiral EFT. Another method utilizing  a unitary transformation in the pion-nucleon Fock
space was used in
Refs.~\cite{Epelbaum:1998ka,Epelbaum:1999dj,Epelbaum:2002gb,Epelbaum:2005fd,Epelbaum:2005bjv,Epelbaum:2007us,Bernard:2007sp,Bernard:2011zr,Krebs:2012yv,Krebs:2013kha,Kolling:2009iq,Kolling:2011mt,Krebs:2016rqz,
  Krebs:2019aka,Krebs:2020plh,Krebs:2020pii} to derive two-, three- and four-nucleon interactions
as well as the electroweak and scalar current operators. Yet another technique based on S-matrix matching was
employed in Refs.~\cite{Kaiser:1997mw,Kaiser:1998wa,Kaiser:1999ff,Kaiser:1999jg,Kaiser:2001dm,Kaiser:2001pc,Kaiser:2001at,Entem:2015xwa} to
calculate the long-distance behavior of two- and three-pion exchange
nucleon-nucleon potentials.
To obtain the unambiguous (up to unitary
transformations) large-distance behavior of the nuclear forces it is
convenient to use dimensional regularization or
equivalent schemes for 
evaluating ultraviolet-divergent loop integrals. In this way, the
expressions for few-nucleon forces in Refs.~\cite{Epelbaum:1999dj,Epelbaum:2002gb,Epelbaum:2005fd,Epelbaum:2005bjv,Epelbaum:2007us,Bernard:2007sp,Bernard:2011zr,Krebs:2012yv,Krebs:2013kha,Kaiser:1997mw,Kaiser:1998wa,Kaiser:1999ff,Kaiser:1999jg,Kaiser:2001dm,Kaiser:2001pc,Kaiser:2001at,Entem:2015xwa} and exchange
currents in Refs.~\cite{Kolling:2009iq,Kolling:2011mt,Krebs:2016rqz,
  Krebs:2019aka,Krebs:2020plh,Krebs:2020pii} have been obtained. On
the other hand, the short-distance asymptotics of the pion-exchange
potentials is dominated by $1/r^n$ singularities, where the power
$n \ge 3$ grows with an increasing order of the chiral EFT
expansion. Clearly, the short-distance or, equivalently,
large-momentum behavior of the pion-exchange potentials
cannot be described in chiral EFT in a meaningful way and needs to be
cut off prior to solving the nuclear Schr\"odinger equation. 
Examples of different types of cutoff regulators in the
chiral two- and three-nucleon forces (3NF) can be found in
Refs.~\cite{Ordonez:1995rz,Epelbaum:1999dj,Epelbaum:2002vt,Epelbaum:2003gr,Epelbaum:2003xx,Epelbaum:2004fk,Entem:2003ft,Gezerlis:2013ipa,Ekstrom:2013kea,Piarulli:2014bda,Epelbaum:2014efa,Epelbaum:2014sza,Ekstrom:2015rta,Entem:2017gor,Reinert:2017usi,LENPIC:2018ewt,Huther:2019ont,Reinert:2020mcu,Maris:2020qne,Somasundaram:2023sup}. 

\begin{figure}[tb]
\vskip 1 true cm
  \begin{center} 
\includegraphics[width=0.9\textwidth,keepaspectratio,angle=0,clip]{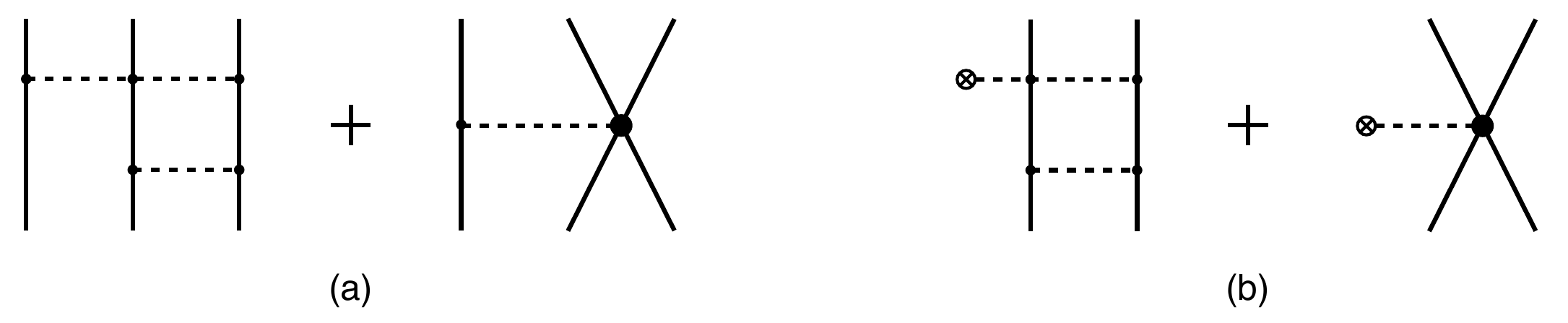}
    \caption{
Examples of contributions to the amplitude for three-nucleon
scattering (a) and for two nucleons interacting with an external axial
source (b) that lead to inconsistencies when using different regularizations in
the potentials and in the Schr\"odinger equation. Solid and dashed lines
refer to nucleons and pions, respectively. Solid dots denote vertices
from the lowest-order pion-nucleon Lagrangian, while filled circles
refer to the $N^\dagger N N^\dagger N \pi$-vertex proportional to the
low-energy constant $c_D$. Circled crosses denote insertions of the external axial source. 
\label{fig1} 
 }
  \end{center}
\end{figure}

On the other hand, it has been shown in Refs.~\cite{Epelbaum:2019kcf,Krebs:2019uvm} that a
simultaneous usage of dimensional regularization for the derivation of nuclear potentials 
and cutoff regularization in the Schr\"odinger equation leads, in
general, to a violation of chiral symmetry, i.e.~it is inconsistent
from the chiral EFT point of view. To see the origin of the problem
consider the contribution to the three-nucleon scattering amplitude
from the first Feynman diagram in Fig.~\ref{fig1} (a). 
When solving the Faddeev
equation, the corresponding contribution emerges from the iteration
of the two-nucleon (2N) one-pion exchange potential $\hat V_{\rm 2N:\, 1\pi}^{(0)}$
and the two-pion exchange 3NF $\hat V_{\rm 3N: \, 2\pi}^{(4)}$
as well as from the genuine 3NF term $\hat V_{\rm 3N: \, 2\pi-1\pi}^{(4)}$, where the superscript
$^{(n)}$ denotes the chiral EFT order $Q^n$, with $Q$ referring to the
small expansion parameter. Both types of
contributions are encoded in the first Feynman diagram of Fig.~\ref{fig1} (a). 
The expressions for the fourth-order (i.e., N$^3$LO) 3NFs  $\hat V_{\rm 3N: \,
  2\pi-1\pi}^{(4)}$ and  $\hat V_{\rm
  3N: \, 2\pi}^{(4)}$ can be found
in Refs.~\cite{Bernard:2007sp} and \cite{Bernard:2011zr}, respectively.
Notice that $\hat V_{\rm 3N: \, 2\pi}^{(4)} \sim {\mathcal{O}} (1/m)$,
where $m$ is the nucleon mass, so that both the iterative
contribution $\hat V_{\rm 3N: \, 2\pi}^{(4)} \, \hat G_0 \, \hat V_{\rm 2N:\,
  1\pi}^{(0)}$, with $\hat G_0$ denoting the three-nucleon resolvent
operator, and the non-iterative term $\hat V_{\rm 3N: \,
  2\pi-1\pi}^{(4)}$
remain finite in the static limit of $m \to
\infty$. The linearly-divergent iterative amplitude $\hat V_{\rm 3N: \, 2\pi}^{(4)} \, \hat G_0 \, \hat V_{\rm 2N:\,
  1\pi}^{(0)}$ has been calculated analytically with a momentum-space
cutoff $\Lambda$ in Ref.~\cite{Epelbaum:2019kcf}. While a part of the resulting
contribution proportional to $\Lambda$ can be absorbed into redefinition
of the low-energy constant $c_D$, see the second diagram in
Fig.~\ref{fig1} (a), the remaining divergent part has the structure that 
violates chiral symmetry and cannot be absorbed by 
counterterms from the effective Lagrangian. Thus,
apparently, renormalization fails for the considered iterative
contribution to the three-nucleon amplitude. As already pointed out,
the problem can be traced back to the mismatch between two different
regularization schemes. Indeed, the problematic divergence would
exactly cancel against the corresponding term stemming from $\hat V_{\rm 3N: \,
  2\pi-1\pi}^{(4)}$ if the latter were calculated using the cutoff
regularization \cite{Epelbaum:2019kcf}. The same issue was shown to affect 
N$^3$LO contributions to the axial current operator in
Ref.~\cite{Krebs:2019uvm}, see Fig.~\ref{fig1} (b). The above discussion
makes it clear that the expressions for the three- and four-nucleon
potentials\footnote{For two-nucleon potentials, the mismatch between
  the different regularization schemes does not lead to 
  inconsistencies as far as the results at the fixed (physical) value
  of the quark masses are concerned. This is because the chiral
  symmetry does not constrain the momentum dependence of the $N^\dagger N N^\dagger
  N$ contact interactions. The $N^\dagger N N^\dagger
  N \pi$ vertices of the $c_D$-type, on the other hand,  must involve derivatives of
  the pion field or quark-mass insertions to comply with the chiral symmetry.} and current operators at N$^3$LO and beyond,
calculated in
Refs.~\cite{Epelbaum:2005bjv,Epelbaum:2007us,Bernard:2007sp,Bernard:2011zr,Krebs:2012yv,Krebs:2013kha,Kolling:2009iq,Kolling:2011mt,Krebs:2016rqz,Krebs:2019aka,Krebs:2020plh,Krebs:2020pii}
using dimensional regularization, need to be re-derived using a cutoff
regulator. This is currently the main obstacle in extending the
precision frontier of chiral EFT studies beyond the purely two-nucleon
sector \cite{Epelbaum:2019kcf,Epelbaum:2022cyo}. 

Clearly, cutoff regularization must be introduced in such a way that the
chiral and gauge symmetries are respected. This can be achieved by
employing a symmetry-preserving regulator at the level of the
effective chiral Lagrangian using, for example, the higher derivative
method introduced by Slavnov  \cite{Slavnov:1971aw}, see also
Refs.~\cite{Djukanovic:2004px,Long:2016vnq}. In the current paper, 
we will, however, consider a simplified model for pion-nucleon interactions and
ignore nontrivial aspects related to preserving
the symmetries of chiral EFT in the presence of a cutoff, which will be addressed in a separate
publication \cite{KE_ToAppear}. Rather, we focus here on developing a method
for deriving nuclear potentials that is sufficiently general to be
applicable to regularized Lagrangians. Specifically, motivated by the
semilocally-regularized 2N potentials of Refs.~\cite{Reinert:2017usi,Reinert:2020mcu}, we
choose the regulator of the pion propagator to be of a Gaussian type.
Notice that in Refs.~\cite{Reinert:2017usi,Reinert:2020mcu}, regularization of the
one- and two-pion exchange potentials was achieved in a somewhat {\it ad
hoc} manner by replacing the static pion propagators with the
regularized ones at the intermediate stage of the calculation. Here,
we will follow a more rigorous approach and introduce the regulator already at the Lagrangian
level in such a way that the modified pion propagator has the form
\beqa
\frac{1}{q_0^2-\vec{q}^{\, 2}-M^2} \; \to \; \frac{e^{\frac{q_0^2-\vec{q}^{\,2}-M^2}{\Lambda^2}}}{q_0^2-\vec{q}^{\,2}-M^2}\,,\label{pion:prop:Reg:Minkowski}
\eeqa 
where $(q^0, \, \vec q \,)$ and $M$ are the pion four-momentum and
mass, respectively, while $\Lambda$ is a cutoff. For static pions with
$q_0 = 0$, the modified pion propagator coincides with 
the one of Ref.~\cite{Reinert:2017usi}. 
The dependence of the regulator in Eq.~(\ref{pion:prop:Reg:Minkowski}) on both the pion
energy $q_0$ and three-momentum $\vec q$ is a consequence of the
Lorentz invariance, which is kept manifest in the pion sector of
chiral EFT. 
We further emphasize that the modification of the pion propagator in Eq.~(\ref{pion:prop:Reg:Minkowski}) does
not affect its long-range behaviour as can be easily verified by 
expanding the right-hand side (rhs) of Eq~(\ref{pion:prop:Reg:Minkowski}) in powers of $1/\Lambda$:
\beqa
\frac{e^{\frac{q_0^2-\vec{q}^{\,2}-M^2}{\Lambda^2}}}{q_0^2-\vec{q}^{\, 2}-M^2}&=&\frac{1}{q_0^2-\vec{q}^{\, 2}-M^2}
+\frac{1}{\Lambda^2}+\frac{1}{2\Lambda^4}
                                                                                                       \left(q_0^2-\vec{q}^{\, 2}-M^2\right)
                                                                                                       +{\cal O}(1/\Lambda^6).\label{SMS:Reg:Expansion}
\eeqa
Except for the original unmodified propagator, all terms on the rhs of
the above equation are polynomial and, therefore, generate only
short-range interactions between nucleons. Stated differently, the
employed regulator does not induce long-range artifacts
(provided $\Lambda$ is chosen sufficiently large).

The modification of the pion propagator as in
Eq.~(\ref{pion:prop:Reg:Minkowski})
can be accomplished by changing the pion kinetic term in the
Lagrangian density via
\beq
\label{LagrRegMinkowski}
-\frac{1}{2} \fet \pi \cdot (\partial_\mu
\partial^\mu + M^2 ) \fet \pi  \;  \to  \; - \frac{1}{2} \fet \pi \cdot (\partial_\mu
\partial^\mu + M^2  )    e^{-\frac{\partial_\mu
\partial^\mu + M^2}{\Lambda^2}} \fet \pi \,.
\eeq
Since we want
to keep the Lorentz symmetry in the pion sector manifest,
i.e.~we refrain from performing the $\partial_0^2/\Lambda^2$-expansion, 
we are facing the complication that the regularized Lagrangian 
involves arbitrarily high-order time derivatives of the pion field.
This makes it difficult to apply Hamiltonian-based
methods that rely on the canonical quantization, 
including  TOPT and the method of unitary transformation, to the derivation
of nuclear forces and currents. For theories with high-order time
derivatives, it is more convenient to employ 
a path-integral formulation, which is commonly used in the
Goldstone boson \cite{Gasser:1983yg} and single-baryon sectors of
ChPT \cite{Jenkins:1990jv,Bernard:1992qa}. 
The purpose of this paper is to describe a path-integral method for 
deriving nuclear forces and current operators
from the regularized chiral Lagrangian. 

\begin{figure}[tb]
\vskip 1 true cm
  \begin{center} 
\includegraphics[width=0.25\textwidth,keepaspectratio,angle=0,clip]{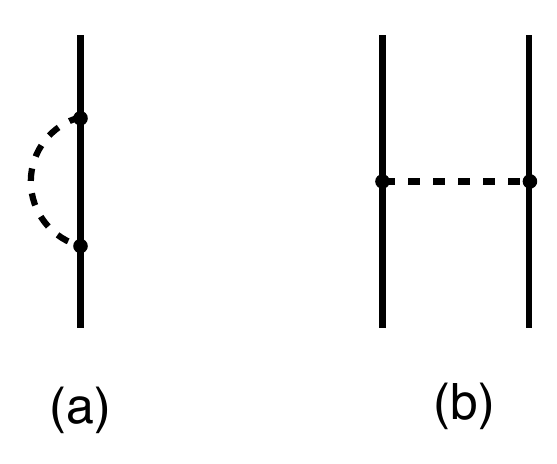}
    \caption{
Nucleon self energy contribution (a) and the one-pion exchange
potential between two nucleons (b).  Solid and dashed lines
refer to nucleons and pions, respectively.
\label{fig2} 
 }
  \end{center}
\end{figure}

Our paper is organized as follows. In sec.~\ref{sec:formalism}, we
introduce a simple toy-model for pion-nucleon scattering by
specifying the corresponding regularized Lagrangian density. After
performing the Euclidean path integral over pion fields we obtain the 
generating functional for $n$-nucleon Green's functions with a
nonlocal action. To determine
nuclear forces, the action is brought to
an instantaneous form by applying (nonlocal) nonlinear redefinitions of the
nucleon fields. We show that this method correctly reproduces the
lowest-order (in the expansion in powers of the coupling constant)
two-nucleon potential due to the one-pion exchange, see
Fig.~\ref{fig2} (b). In sec.~\ref{sec:Determinant}, we
describe how loop contributions to the nuclear Hamiltonian are generated
by the functional determinant stemming from the nucleon field
redefinitions. As an example, we consider the self-energy diagram and
calculate the leading one-loop contribution to the nucleon mass, see
Fig.~\ref{fig2} (a), using our new approach. Finally, in sec.~\ref{sec:ZFactor}, we discuss the
dependence of  the nucleon $Z$-factor on the choice of external
sources.
The main results of this paper are summarized in
sec.~\ref{sec:summary}, while appendices \ref{functional:det} and \ref{zetafunctionReg}
provide further technical details related to the calculation of the
functional determinant.

\section{Instantaneous nuclear interactions via nonlinear field redefinitions}
\def\theequation{\arabic{section}.\arabic{equation}}
\label{sec:formalism}

The toy-model we consider here consists of light pseudoscalar pions
interacting with non-relativistic nucleons via a pseudovector
coupling.   The Lagrangian density of the toy-model is given by
\beqa
{\cal L}&=&N^\dagger\bigg(i\frac{\partial}{\partial
  x_0}+\frac{\vec{\nabla}^{\,2}}{2 m}+\frac{g}{2 F} \,\vec{\sigma}\cdot\vec{\nabla}\fet{\pi}\cdot\fet{\tau}\bigg)N +
  \frac{1}{2}\left(\partial_\mu\fet{\pi}\cdot\partial^\mu\fet{\pi} -M^2\fet{\pi}^2\right).
\eeqa
The constants $M, g, F$ and $m$ denote the pion mass, axial
coupling constant, pion decay constant and the nucleon mass, respectively.
The derivative-coupling of the pion field $\fet{\pi}$ to the nucleon field
$N$ is the same as in ChPT at lowest order, but we left out all
other pion-nucleon and four-pion vertices for the sake of simplicity.
Since we are interested in nuclear potentials for
three-momenta of the order of $M \ll m$, we treat nucleons
non-relativistically but keep the fully relativistic treatment of pion
dynamics. Further, $\sigma^a$ and $\tau^a$ with
$a=1,2,3$ denote Pauli spin and isospin matrices, respectively. The generating
functional is given by
\beqa
\label{ZMink}
Z[\eta^\dagger, \eta]&=&\int DN^\dagger DN D\fet\pi
\exp\left(iS+i\int d^4 x\left[\eta^\dagger N+N^\dagger\eta\right]\right),
\eeqa
with the action
\beqa
S&=&\int d^4 x \,{\cal L}.
\eeqa
Clearly, $n$-nucleon Green's functions can be extracted from $Z[\eta^\dagger, \eta]$ by
taking functional derivatives in the Grassmann variables $\eta^\dagger$
and $\eta$ referring to the external sources and setting them to zero.
Note that we do not introduce external sources attached to
the pion since we restrict ourselves to the kinematics below the
pion-production threshold. The pion field can thus
be considered as an auxiliary field that only appears inside the
path-integral in Eq.~(\ref{ZMink}).

As explained in sec.~\ref{sec:Intro}, we intend to regularize the
theory by modifying the pion kinetic term in the Lagrangian as given
in Eq.~(\ref{LagrRegMinkowski}). However, in Minkowski space, the
above modification obviously does not act as a regulator since the modified
propagator grows exponentially with $q_0$.
For this reason, we will switch from now on to the formulation
in Euclidean space and perform a Wick rotation back to Minkowski
space at the end of the calculation. The generating functional in
Euclidean space is obtained from Eq.~(\ref{ZMink}) via a Wick rotation
by formally replacing $x_0\to -ix_0$,
\beqa
Z[\eta^\dagger,\eta]&=&\int DN^\dagger DN D\fet{\pi}
\exp\left(-\,S^{\rm E}+\int
  d^4x \,\left[\eta^\dagger N+N^\dagger\eta\right]\right),\label{Generating:Functional:Eucl}
\eeqa
where the Euclidean action is given by 
\beqa
\label{ActionNoReg}
S^{\rm E}&=&\int d^4 x\, {\cal L}^{\rm E},\quad \quad
{\cal L}^{\rm E}\,=\,N^\dagger\left(\frac{\partial}{\partial
  x_0}-\frac{\vec{\nabla}^{\,2}}{2 m}-\frac{g}{2 F}\vec{\sigma}\cdot\vec{\nabla} \,\fet{\pi}\cdot\fet{\tau}\right)N +
  \frac{1}{2}\left(\partial_\mu\fet{\pi}\cdot\partial_\mu\fet{\pi} +M^2\fet{\pi}^2\right).
\eeqa
To obtain the expression for the nucleon propagator in Euclidean space
consider a non-interacting generating functional
\beqa
Z_0[\eta^\dagger,\eta]&=&\int DN^\dagger DN \exp\bigg(-\int d^4 x N^\dagger\bigg(\frac{\partial}{\partial
  x_0}-\frac{\vec{\nabla}^{\,2}}{2 m}\bigg)N+\int
  d^4x \,\Big[\eta^\dagger N+N^\dagger\eta\Big]\bigg)\nn
&=& \alpha \exp\bigg(\int d^4 x_1 d^4 x_2\eta^\dagger(x_1) P^{\rm E}(x_1-x_2)\eta(x_2)\bigg)\,,
\eeqa
with $\alpha$ some normalization constant and the
propagator $P^{\rm E}(x_1-x_2)$ representing the solution to the
equation 
\beqa
\bigg(\frac{\partial}{\partial x_1^0} - \frac{\vec{\nabla}_{x_1}^2}{2
  m}\bigg) P^{\rm E}(x_1-x_2)&=&\delta^4 (x_1-x_2)\,,\label{FreeEuclideanNucleonPropagator}
\eeqa
subject to the boundary condition
\beqa
\lim_{x_0\to\infty} P^{\rm E} (x)&=&0\,.
\eeqa
Its explicit expression has the form 
\beqa
P^{\rm E}(x)&=&\int\frac{d^4
  p}{(2\pi)^4}\frac{1}{ip_0+\frac{\vec{p}^{\, 2}}{2m}+\epsilon} e^{i (p_0 x_0+\vec{p}\cdot\vec{x})}\,.
\eeqa
Since the pole-term is in the upper half of the complex $p_0$-plane, we get even in the
static limit an exponential decrease with increasing positive $x_0$ as
long as $\epsilon > 0$,
\beqa
\lim_{m\to\infty}P^{\rm E}(x)=\theta(x_0)\delta(\vec{x})\, e^{-\epsilon\,x_0}\,.
\eeqa

The regularized version of the Lagrangian given in Eq.~(\ref{ActionNoReg}) has the form
\beqa
{\cal L}^{\rm E}_\Lambda&=&N^\dagger\bigg[\frac{\partial}{\partial
  x_0}\bigg(1+\frac{1}{\Lambda_\tau}\frac{\partial}{\partial x_0}\bigg)-\frac{\vec{\nabla}^{\,2}}{2 m}-\frac{g}{2 F}\vec{\sigma}\cdot\vec{\nabla} \,\fet{\pi}\cdot\fet{\tau}\bigg]N +
  \frac{1}{2}\fet{\pi}\cdot\left(-\partial^2 +M^2\right)\exp\left(\frac{-\partial^2+M^2}{\Lambda^2}\right)\fet{\pi},\label{Regularized:LE}
\eeqa
where $\partial^2 \equiv \partial_\mu^2$. Note that in addition to the
modification of the pion kinetic term already discussed in
sec.~\ref{sec:Intro}, we have also introduced a temporal regulator in the
nucleon propagator. The need for such a regulator and its explicit choice,
which ensures that the dynamical equation has the form of the
Schr\"odinger or Lippmann-Schwinger equation,
will be justified below. Notice that at the end of the calculation, we will
take the limit $\Lambda_\tau \to \infty$ (while keeping the cutoff
$\Lambda$ finite), so that our results will turn out to be independent of
$\Lambda_\tau$.

Starting from Eq.~(\ref{Generating:Functional:Eucl}) with the
regularized Lagrangian given in Eq.~(\ref{Regularized:LE}), the
integration over the pion field can be trivially performed since
the Lagrangian density ${\cal L}^{\rm E}_\Lambda$ is a quadratic function in
$\fet \pi$. After integrating out the pion field, we obtain 
\beq
Z[\eta^\dagger, \eta]= A \int DN^\dagger DN
\exp\left(-S_N^{\rm E}+\int d^4x \left[\eta^\dagger
    N+N^\dagger\eta\right]\right),\label{path:int:eucl}
\eeq
where the action now depends only on the nucleon fields:
\beqa
S_N^{\rm E}&=&\int d^4 x N^\dagger\bigg[\frac{\partial}{\partial
  x_0}\left(1+\frac{1}{\Lambda_\tau}\frac{\partial}{\partial x_0}\right)-\frac{\vec{\nabla}^{\,2}}{2 m}\bigg]N\nn
&-&\frac{g^2}{8 F^2}\int d^4x_1 \,d^4x_2 \,
\vec{\nabla}_{x_1}\cdot\left[N^\dagger(x_1)\vec{\sigma}\,\fet{\tau}N(x_1)\right]\Delta^{\rm
E}(x_1-x_2)\vec{\nabla}_{x_2}\cdot\left[N^\dagger(x_2)\vec{\sigma}\,\fet{\tau}N(x_2)\right] ,\quad\label{NN:NonInstant}
\eeqa
with the Euclidean propagator
\beq
\Delta^{\rm E}(x) = \int \frac{d^4q}{(2\pi)^4} e^{i q\cdot x} \, \frac{e^{-\frac{q_0^2+\vec{q}^{\, 2}+M^2}{\Lambda^2}}}{q_0^2+\vec{q}^{\, 2}+M^2}.
\eeq
The obtained result is still not very useful for the purpose of
deriving nuclear forces. Indeed, while the lowest-order (i.e., order-$g^2$) connected contribution to
the $4$-point function matches the regularized one-pion exchange
potential\footnote{Note the prefactor $1/8$ in front of the
  one-pion-exchange contribution. To obtain the expression
  corresponding to the Feynman
  diagram (b) in Fig.~\ref{fig2} from Eq.~(\ref{NN:NonInstant}) one has to
  calculate the
  functional derivatives of the action $S_N^{\rm E}$ with respect to the nucleon
  fields. Because of two nucleon lines, one obtains a factor of $2$,
  so that the one-pion exchange potential acquires the correct
  prefactor $- g^2/(4F^2)$.}, higher-order
contributions do not immediately have the form
of iterations of the Lippmann-Schwinger equation. Thus,
nuclear potentials at orders beyond $\mathcal{O} (g^2)$ cannot be directly read off from
Eq.~(\ref{NN:NonInstant}). To be able to interpret the interaction
part of the action $S_N^{\rm E}$  as the integral of the nuclear Hamiltonian density, the
action needs to be made instantaneous, i.e., all pion propagators
$\Delta^{\rm E}(x)$ have to be replaced by the static ones
\beqa
\Delta^{\rm S}(x)&=&\delta(x_0)\bar{\Delta}^{\rm S}(\vec{x}\,), \quad \quad
\bar{\Delta}^{\rm S}(\vec{x}\,)\,=\,\int \frac{d^3q}{(2\pi)^3} e^{i
  \vec{q}\cdot \vec{x}} \, 
\frac{e^{-\frac{\vec{q}^{\, 2}+M^2}{\Lambda^2}}}{\vec{q}^{\, 2}+M^2}.
\label{DeltaSDefinition}
\eeqa
As will be shown below, this can be achieved by performing
nonlinear (and nonlocal) redefinitions of the nucleon field, which
leave S-matrix elements unchanged, see also Ref.~\cite{Gasparyan:2021edy} for an alternative
method. A similar technique has  already been
used in the context of nuclear chiral EFT to facilitate the derivation
of isospin-breaking nuclear potentials by eliminating the single-nucleon vertex
proportional to the neutron-proton mass difference \cite{Friar:2004ca,Epelbaum:2007sq}. 
A path-integral formulation with an instantaneous action was also 
discussed within lattice EFT in Ref.~\cite{Borasoy:2006qn}.

Before proceeding with the calculations, it is instructive to address the
role of the temporal regulator in Eq.~(\ref{NN:NonInstant}). Without
such a regulator, the replacement of the pion propagator
$\Delta^{\rm E}(x)$ by its static version $\Delta^{\rm S}(x)$ would lead to 
divergent loop integrals. To be specific, consider the one-loop
self-energy diagram (a) in Fig.~\ref{fig2} obtained from the
action $S_N^{\rm E}$ with the temporal regulator being switched off by taking
the limit $\Lambda_\tau \to \infty$. The corresponding loop integral
in the static limit has the form
\beqa
\Sigma_{\rm E} (q_0)&=&\int \frac{d^4
  l}{(2\pi)^4}\frac{1}{i(l_0+q_0)+\epsilon}\, \frac{e^{-\frac{l_0^2+\vec{l}^{\, 2}+M^2}{\Lambda^2}}}{l_0^2+\vec{l}^{\, 2}+M^2}.
\eeqa
Suppose now that we have succeeded to achieve our goal and replaced
all non-instantaneous interactions appearing in the path-integral by the instantaneous
ones. Then, the corresponding self-energy integral would take the form
\beqa
\Sigma_{\rm S} (q_0)&=&\int \frac{d^4
  l}{(2\pi)^4}\frac{1}{i(l_0+q_0)+\epsilon} \, \frac{e^{-\frac{\vec
      l^{\, 2}+M^2}{\Lambda^2}}}{\vec l^{\, 2}+M^2},\label{SigmaSUnregularized}
\eeqa
which is not well-defined since the $l_0$-integration diverges
logarithmically. We can cure this by multiplying the nucleon propagator
with the temporal regulator such that
Eq.~(\ref{SigmaSUnregularized}) turns to
\beqa
\Sigma_{\rm S}(q_0)&=&\int \frac{d^4
  l}{(2\pi)^4}\frac{1}{i(l_0+q_0)+\epsilon}\frac{\Lambda_\tau}{i
  (l_0+q_0)+\Lambda_\tau} \, \frac{e^{-\frac{\vec l^{\, 2}+M^2}{\Lambda^2}}}{\vec l^{\, 2}+M^2}\;=\;0.\label{SigmaSReegularized}
\eeqa
The last equality holds by virtue of the residue theorem. Notice that
the vanishing of the self-energy contribution obtained from the generating
functional in Eq.~(\ref{path:int:eucl}) is precisely what one would expect
if the action $S_N^{\rm E}$ involves only instantaneous  pion exchange
interactions. This also restricts possible choices of the temporal
regulator. For example, a symmetric (in the derivative operator)
choice like
\beqa
\int d^4 x \, N^\dagger (x) \frac{\partial}{\partial
  x_0} N (x) &\to &
\int d^4 x d^4 y \, N^\dagger (x) \bigg[\frac{\partial}{\partial
  x_0}{\big(R^{\Lambda_\tau}\big)}^{-1}(x-y)\bigg]N (y) 
\nonumber
\eeqa
with
\beq 
R^{\Lambda_\tau}(x-y)=\frac{\Lambda_\tau^2}{-[\partial_x^0]^2+\Lambda_\tau^2}
\, \delta^4 (x-y), \quad \quad{\rm and}\quad \quad \big( R^{\Lambda_\tau}\big)^{-1} (x-y)=
\frac{-[\partial_x^0]^2+\Lambda_\tau^2 }{\Lambda_\tau^2}\delta^4(x-y)\,,
\eeq
would lead to a non-vanishing result for the self-energy contribution
\beqa
\Sigma_{\rm S} (q_0)&=&\int \frac{d^4
  l}{(2\pi)^4}\, \frac{1}{i(l_0+q_0)+\epsilon}\, \frac{\Lambda_\tau^2}{
  (l_0+q_0)^2+\Lambda_\tau^2}\, \frac{e^{-\frac{\vec l^{\, 2}+M^2}{\Lambda^2}}}{\vec l^{\, 2}+M^2}\nn
&=&\frac{1}{2}\int \frac{d^3 l}{(2\pi)^3}\, \frac{e^{-\frac{\vec l^{\, 2}+M^2}{\Lambda^2}}}{\vec l^{\, 2}+M^2}\,\neq\,0.
\eeqa
To avoid this complication, we choose the temporal regulator to have all poles in
the upper half of the complex $l_0$-plane. Our choice of
the temporal regulator corresponds to Eq.~(\ref{NN:NonInstant}):
\beqa
R^{\Lambda_\tau}(x-y)&=&\frac{\Lambda_\tau}{\partial_x^0+\Lambda_\tau}\,
\delta^4(x-y)\;=\;\theta(x_0-y_0)\delta(\vec{x}-\vec{y}\,)\Lambda_\tau e^{-\Lambda_\tau(x_0-y_0)}.\label{temporalRegulatorDefinition} 
\eeqa
Note that this regulator satisfies the proper boundary condition $\lim_{x_0\to\infty}R^{\Lambda_\tau}(x)=0$.

After these preparations, we now focus on the
modification of the Euclidean action, aiming to replace all
non-instantaneous interactions by instantaneous ones. We start
with a trivial identity 
\beqa
\Delta^{\rm E}(x)&=&\Delta^{\rm S}(x)+\Delta^{\rm E}(x)-\Delta^{\rm S}(x).\label{DiffEq}
\eeqa
The difference $\Delta^{\rm E}(x)-\Delta^{\rm S}(x)$ can be written in the form
\beqa
\Delta^{\rm E}(x)-\Delta^{\rm S}(x)&=&\frac{\partial^2}{\partial
  x_0^2}\Delta^{\rm ES}(x),
\eeqa
where
\beqa
\Delta^{\rm ES}(x)&=&-\int \frac{d^4q}{(2\pi)^4} e^{i q\cdot x} \frac{1}{q_0^2}\Bigg(\frac{e^{-\frac{q_0^2+\vec{q}^{\, 2}+M^2}{\Lambda^2}}}{q_0^2+\vec{q}^{\, 2}+M^2}-\frac{e^{-\frac{\vec{q}^{\, 2}+M^2}{\Lambda^2}}}{\vec{q}^{\, 2}+M^2}\Bigg).\label{DeltaES:Def}
\eeqa
Notice that the integrand in Eq.~(\ref{DeltaES:Def}) is a smooth
function at $q_0=0$.  In the limit $\Lambda\to\infty$,
it has a simple form
\beqa
\lim_{\Lambda\to\infty} \Delta^{\rm ES}(x)&=&\int
\frac{d^4q}{(2\pi)^4} e^{i q\cdot x}\frac{1}{q_0^2+\vec{q}^{\, 2}+M^2}\frac{1}{\vec{q}^{\, 2}+M^2},
\eeqa
which, in momentum space, reduces to a product of the Euclidean and static pion
propagators.  This motivates the superscript 
ES of the propagator in Eq.~(\ref{DeltaES:Def}). Thus, the
trivial identity in Eq.~(\ref{DiffEq}) is rewritten as
\beqa
\Delta^{\rm E}(x)&=&\Delta^{\rm S}(x)+\frac{\partial^2}{\partial
  x_0^2}\Delta^{\rm ES}(x).\label{Diff:Modified}
\eeqa
Let us now replace the Euclidean non-static pion propagator in
Eq.~(\ref{NN:NonInstant}) by the rhs of Eq.~(\ref{Diff:Modified}). The
first term on rhs of Eq.~(\ref{Diff:Modified}) already generates an instantaneous 
one-pion exchange interaction. The second term 
\beqa
&-&\frac{g^2}{8 F^2}\int d^4x_1 \,d^4x_2 \,
\vec{\nabla}_{x_1}\cdot\left[N^\dagger(x_1)\vec{\sigma}\,\fet{\tau}N(x_1)\right]\frac{\partial^2}{\partial
  {x_1^0}^2}\Delta^{\rm ES}(x_1-x_2)
\vec{\nabla}_{x_2}\cdot\left[N^\dagger(x_2)\vec{\sigma}\,\fet{\tau}N(x_2)\right]
\nn
&&\quad {} = \;\frac{g^2}{8 F^2}\int d^4x_1 \,d^4x_2 \,
N^\dagger(x_1)\vec{\sigma}\,\fet{\tau}N(x_1)\cdot\frac{\partial^2}{\partial
  {x_1^0}^2}\vec{\nabla}_{x_1}\otimes\vec{\nabla}_{x_1}\Delta^{\rm ES}(x_1-x_2)
\cdot N^\dagger(x_2)\vec{\sigma}\,\fet{\tau}N(x_2)\label{termtoeliminate}
\eeqa
involves
time derivatives
and thus can be eliminated via
a suitable nucleon field redefinition. In contrast to the usually
considered field transformations, see e.g.~\cite{Friar:2004ca}, the field
redefinitions required here are
nonlocal.

Before specifying the field redefinitions, it is useful to introduce a
compact notation for writing nonlocal interactions. As one can
see already from Eq.~(\ref{termtoeliminate}), there appear a number of
integrations over coordinates $x_i$, and one needs to use the tensor language. It is advantageous
to employ a shortcut notation  to keep the presentation more
transparent. We introduce the following abbreviations:
\beqa
\int d^4 x_i \,d^4 x_j\,N^\dagger(x_i) \Gamma N(x_i) \Delta^{\rm X}(x_i-x_j) 
N^\dagger(x_j) \Gamma ' N(x_j) \; \to \;  N^\dagger_i  \Gamma N_i \, 
\Delta^{\rm X}_{ij} \, N^\dagger_j \Gamma '  N_j,\nonumber
\eeqa
where $\Gamma$, $\Gamma '$  are some spin-isospin operators and the
superscript X stands for E,
S or ES. Here and in what follows, we drop the integral symbols, but the  integration
over all coordinates is to be
understood. The time and
spatial derivatives are abbreviated as
\beq
\frac{\partial}{\partial x_i^0} N(x_i) \; \to \; \partial_i^0
N_i,\quad \quad
\vec{\nabla}_{x_i} N(x_i)\;\to\;\vec{\nabla}_i N_i\,,
\eeq
while the propagators are abbreviated as
\beq
X(x_1-x_2)\to X_{12}, \quad\quad  X = R^{\Lambda_\tau},\;
(R^{\Lambda_\tau})^{-1}, \; \Delta^{{\rm E}}, \;\Delta^{{\rm S}},
\; \Delta^{{\rm ES}}\,.
\eeq
For spin and isospin matrices, we use
\beq
\vec{\sigma}\fet{\tau}N(i) \; \to  \; \vec{\sigma}_i\fet{\tau}_i N_i
\; \equiv \; N_i\vec{\sigma}_i\fet{\tau}_i\,.
\eeq
The last equality signifies that we do not explicitly keep the proper
ordering of the Pauli matrices and the corresponding spinors in order
to simplify the notation. 
Finally, we will also often suppress the structures like
$N^\dagger_iN_i^{}$ for brevity while assuming them to
be there. Only the structures $N^\dagger_i\partial_i^0N_i$, 
$N^\dagger_i\vec{\nabla}_i^2N_i$ and single appearances of the
nucleon field will always be written out explicitly.
Using this shortcut notation,
the action in Eq.~(\ref{NN:NonInstant}) is written in a compact form as
\beqa
S_N^{\rm E}&=& N^\dagger_1 \partial_1^0
  \big(R^{\Lambda_\tau}\big)_{12}^{-1}N_2-N^\dagger_1 \frac{\vec{\nabla}_1^{2}}{2
    m}N_1+\frac{g^2}{8 F^2}
\fet{\tau}_1\cdot\fet{\tau}_2\, \big(
\vec{\sigma}_1\cdot\vec{\nabla}_1 \vec{\sigma}_2\cdot\vec{\nabla}_1 \Delta^{\rm
  E }_{12} \big)\nn
&=&N^\dagger_1\partial_1^0
  \big(R^{\Lambda_\tau}\big)_{12}^{-1}N_2-N^\dagger_1\frac{\vec{\nabla}_1^{2}}{2
    m}N_1  +\frac{g^2}{8 F^2}
  \fet{\tau}_1\cdot\fet{\tau}_2\, 
\Big\{\big(\vec{\sigma}_1\cdot\vec{\nabla}_1 \vec{\sigma}_2\cdot\vec{\nabla}_1 
  \Delta^{\rm
  S}_{12} \big)
+ \Big[ \vec{\sigma}_1\cdot\vec{\nabla}_1 \vec{\sigma}_2\cdot\vec{\nabla}_1   (\partial_1^0)^2\Delta^{\rm
  ES}_{12}\Big] \Big\},\label{euclactiong2:compact}
\eeqa
where the temporal regulator is defined in Eq.~(\ref{temporalRegulatorDefinition}).

To get an idea of how one can eliminate the non-instantaneous term in
the last line of
Eq.~(\ref{euclactiong2:compact}) involving $\Delta^{\rm
  ES}_{12}$ consider a redefinition of the nucleon field via
\beqa
N_1&\to&N_1 + \Gamma R^{\Lambda_\tau}_{12} N_2  f_{23}  (N^\dagger_3
\Gamma ' N_3),\nn
N^\dagger_1&\to&N^\dagger_1+(N^\dagger_3\Gamma'^\dagger N_3) 
f^\star_{32}  N^\dagger_2 R^{\Lambda_\tau}_{21} \Gamma^\dagger
,
\eeqa
with $f_{ij} \equiv f(x_i-x_j)$, $f^\star_{ij} \equiv [f
(x_i-x_j)]^\star$ being some smooth functions and $\Gamma$, $\Gamma '$ spin-isospin operators.
We remind the reader that the integration over the coordinates $x_2$
and $x_3$, which is not
written out explicitly, is to be understood. 
The part of the kinetic term with the time derivative
in Eq.~(\ref{euclactiong2:compact}) transforms to
\beqa
N^\dagger_1 \partial_1^0 \big(R^{\Lambda_\tau}\big)^{-1}_{12}N_2&\to&
\Big[N^\dagger_1+(N^\dagger_3\Gamma'^\dagger N_3 ) 
f^\star_{32}  N^\dagger_2 R^{\Lambda_\tau}_{21} \Gamma^\dagger \Big] \;
\partial_1^0 \big(R^{\Lambda_\tau}\big)^{-1}_{14} \,
\Big[N_4 + \Gamma R^{\Lambda_\tau}_{45} N_5  f_{56}  ( N^\dagger_6
\Gamma ' N_6) \Big]
\nn
&=& N^\dagger_1 \partial_1^0 \big( R^{\Lambda_\tau}_{12} \big)^{-1}N_2 \nn
& + &  N^\dagger_1  \Gamma  N_1
   \partial_1^0 f_{12}\big( N^\dagger_2 \Gamma ' N_2 \big)\;+\; (N^\dagger_1  \Gamma \partial_1^0  N_1)
   f_{12}\big( N^\dagger_2 \Gamma ' N_2 \big)
\; + \; (N^\dagger_2\,\Gamma'^\dagger N_2 )  f^\star_{21}  ( N^\dagger_1 \Gamma^\dagger\,\partial_1^0 N_1)
 \nn
&+&   
  (N^\dagger_3\,\Gamma'^\dagger\, N_3)  f^\star_{31} ( N^\dagger_1
  \Gamma^\dagger \Gamma \partial_1^0 R^{\Lambda_\tau}_{12} N_2 f_{24} )
  ( N^\dagger_4\,\Gamma' N_4). \label{fieldredef}
  \eeqa
The second and third lines of the rhs of Eq.~(\ref{fieldredef}) comprise two-
and three-nucleon interactions, respectively. Whenever a
nucleon field
redefinition is performed, new interactions involving time derivatives of nucleon
fields are generated. However, for functions $f$ that fulfill 
$f^\star(x)=-f(-x)$ and for Hermitian operators $\Gamma = \Gamma^\dagger$,
$\Gamma' = \Gamma '^\dagger$,
the time derivative of the 
nucleon fields cancels out in the two-nucleon interactions, and we obtain
\beqa
N^\dagger_1 \partial_1^0 \big(R^{\Lambda_\tau}_1\big)^{-1}N_1 & \to
& N^\dagger_1 \partial_1^0 \big( R^{\Lambda_\tau}_1 \big)^{-1}N_1 \nn
&+&  (N^\dagger_1  \Gamma N_1 \big)
  \big( \partial_1^0 f_{12} \big) (N^\dagger_2 \Gamma ' N_2) \nn
&-&  
 (N^\dagger_3\Gamma' N_3) \, f_{13} \,( N^\dagger_1
  \Gamma^2 \partial_1^0 R^{\Lambda_\tau}_{12} N_2 f_{24} )
 \, ( N^\dagger_4\Gamma' N_4) . \label{fieldredefmod}
 \eeqa
We now  apply this result to Eq.~(\ref{euclactiong2:compact}). To
eliminate the non-instantaneous term in
Eq.~(\ref{euclactiong2:compact}) with the propagator $\Delta_{12}^{\rm
ES}$, we choose the field redefinitions of the form 
\beqa
N_1&\to&N_1-\frac{g^2}{8 F^2}\, R^{\Lambda_\tau}_{13} N_3
\, \fet{\tau}_3\cdot\fet{\tau}_2  \, \big(
\vec{\sigma}_3\cdot\vec{\nabla}_3  \vec{\sigma}_2\cdot\vec{\nabla}_3 \partial_3^0\Delta^{\rm
  ES}_{32} \big),\nn
N^\dagger_1&\to&N^\dagger_1+\frac{g^2}{8 F^2}
\, \big(\vec{\sigma}_3\cdot\vec{\nabla}_3 \vec{\sigma}_2\cdot\vec{\nabla}_3
\partial_3^0\Delta^{\rm
  ES}_{32} \big)\, \fet{\tau}_3\cdot\fet{\tau}_2 \, N^\dagger_3 R^{\Lambda_\tau}_{31}.\label{ope:field:redef}
\eeqa
Obviously, this kind of transformations will generate, apart from the
static two-nucleon interactions, also relativistic
corrections and three-nucleon interactions. If we neglect
relativistic contributions and terms of order $\mathcal{O} (g^{n})$
with $n > 2$, 
we obtain for the transformed
Euclidean action:
\beqa
S_N^{\rm E}&\to&N^\dagger_1\partial_1^0
\big(R^{\Lambda _\tau}\big)^{-1}_{12}N_2 -N^\dagger_1\frac{\vec{\nabla}_1^{\,2}}{2 m}N_1 \;
+ \; \frac{g^2}{8 F^2} 
\fet{\tau}_1\cdot\fet{\tau}_2 \, \big( \vec{\sigma}_1\cdot\vec{\nabla}_1 \vec{\sigma}_2\cdot\vec{\nabla}_1\Delta^{\rm
  S}_{12} \big)\; 
+\;  {\cal O}\bigg(g^4,\,  \frac{g^2}{m}\bigg).\label{euclactiong2}
\eeqa

It is important to emphasize that the
contributions $\eta^\dagger N$ and $N^\dagger \eta$ in
Eq.~(\ref{path:int:eucl}) also change upon
performing the nucleon field redefinitions in Eq.~(\ref{ope:field:redef}). However, the
induced additional terms
\beq
-\frac{g^2}{8 F^2}
\eta^\dagger_1
R^{\Lambda_\tau}_{13} N_3\,
\fet{\tau}_3\cdot\fet{\tau}_2\,
\big(\vec{\sigma}_3\cdot\vec{\nabla}_3 \vec{\sigma}_2\cdot\vec{\nabla}_3 \partial_3^0\Delta^{\rm
  ES }_{32}\big) \; + \; \frac{g^2}{8 F^2}
N^\dagger_3 R^{\Lambda_\tau}_{31}\eta_1 \, 
\fet{\tau}_3\cdot\fet{\tau}_2\,
\big(\vec{\sigma}_3\cdot\vec{\nabla}_3 \vec{\sigma}_2\cdot\vec{\nabla}_3\partial_3^0\Delta^{\rm
  ES}_{32} \big),\label{eta:higher:N}
\eeq
do not produce on-shell singularities in the Green's functions
associated with 
external nucleon legs. Thus, using the LSZ formalism to obtain
S-matrix elements by multiplying the $n$-point functions
with the inverse nucleon propagators and going on-shell, the results
for observables will be unaffected by dropping the
contributions to the generating functional from the terms 
in Eq.~(\ref{eta:higher:N})
in accordance with the equivalence theorem in quantum field
theory \cite{Haag:1958vt,Coleman:1969sm}. We will come back to the role of these terms in sec.~\ref{sec:ZFactor}. 

One final remark is in order regarding the choice of the field redefinitions in
Euclidean space. It is well known that the Euclidean field
$N^\dagger$ is not a complex conjugate of the Euclidean $N$-field. For this
reason, field transformations of $N$ and
$N^\dagger$ fields can, in principle, be chosen independent of each
other. We, however, further constrain them by requiring that
the resulting nuclear interactions are Hermitian after performing the
Wick rotation back to Minkowski
space. To elaborate on this point, it is instructive to look at the
field redefinitions we would apply in Minkowski space if
the convergence of the Gaussian regulator were not an issue.
To find the corresponding Minkowski-space field redefinitions we 
replace in the Euclidean-space expressions every $0$th
component of a four-vector by $i$ times the same component,
e.g.~$x_0\to i x_0$, $q_0\to i q_0$, etc. Notice that this way,
$\Delta^{\rm ES}$ turns into the Feynman-static propagator 
$\Delta^{\rm ES}(x)\to i \Delta^{\rm FS}(x)$ and $R^{\Lambda_\tau}(x) \to iR^{\Lambda_\tau , \rm M}(x)$, where
\beqa
\Delta^{\rm FS}(x)&=&\int \frac{d^4q}{(2\pi)^4} e^{i (-q_0
  x_0+\vec{q}\cdot\vec{x})}
\frac{1}{q_0^2}\Bigg(\frac{e^{-\frac{-q_0^2+\vec{q}^{\,
        2}+M^2}{\Lambda^2}}}{-q_0^2+\vec{q}^{\,
    2}+M^2}-\frac{e^{-\frac{\vec{q}^{\,
        2}+M^2}{\Lambda^2}}}{\vec{q}^{\, 2}+M^2}\Bigg)\,,\nn
R^{\Lambda_\tau , \rm M}(x)&=&\int\frac{d^4
  p}{(2\pi)^4}\frac{-\Lambda_\tau}{p_0-\Lambda_\tau+i\epsilon}\, e^{i(-p_0
  x_0+\vec{p}\cdot\vec{x})}.\label{FS:Propagator}
\eeqa
The first of the above expressions is obviously not well-defined due
 to the exponential increase with growing $q_0$, which is why
 we performed the calculations in Euclidean space.\footnote{With a different
 regulator like, e.g., the Pauli-Villars one, the expression for
 $\Delta^{\rm FS}(x)$ would be well-defined.}
With these preparations, we obtain the Minkowski-space version of the
field redefinitions in Eq.~(\ref{ope:field:redef}):
\beqa
N_1&\to&N_1-\,i\frac{g^2}{8 F^2}\, R^{\Lambda_\tau, \rm M}_{13} N_3
\, \fet{\tau}_3\cdot\fet{\tau}_2  \, \big(
\vec{\sigma}_3\cdot\vec{\nabla}_3  \vec{\sigma}_2\cdot\vec{\nabla}_3 \partial_3^0\Delta^{\rm
  FS}_{32} \big),\nn
N^\dagger_1&\to&N^\dagger_1+\,i\frac{g^2}{8 F^2}
\, \big(\vec{\sigma}_3\cdot\vec{\nabla}_3 \vec{\sigma}_2\cdot\vec{\nabla}_3
\partial_3^0\Delta^{\rm
  FS}_{32} \big)\, \fet{\tau}_3\cdot\fet{\tau}_2 \, N^\dagger_3
R^{\Lambda_\tau, \rm M}_{31}. \label{ope:field:redef:minkowski}
\eeqa
Using Eq.~(\ref{ope:field:redef:minkowski}), one immediately verifies that the
transformed $N^\dagger$-field is a complex conjugate of the
transformed $N$-field, which guarantees that the obtained nuclear
potentials are real-valued. Every field redefinition we apply is
required to fulfill this condition.

Once all interactions in the action are brought to the instantaneous
form, one can perform the Wick rotation back to Minkowski
space. Obviously, this is only possible if the incoming and outgoing
energies of the nuclear system under consideration are below the pion-production
threshold, since otherwise one would hit the cut and the 
potentials would acquire an imaginary part.
After the Wick rotation, the Euclidean action $S_N^{\rm E}$
in Eq.~(\ref{euclactiong2}) turns to $-i S_N$ with
\beq
S_N \; = \;  N^\dagger_1 i \,\partial_1^0 \big(R^{\Lambda_\tau , \rm
  M}\big)^{-1}_{12} N_2
+N_1^\dagger\frac{\vec{\nabla}_1^{\,2}}{2 m}N_1 \; 
-\;  \frac{g^2}{8 F^2}\fet{\tau}_1\cdot\fet{\tau}_2
\,
\big(\vec{\sigma}_1\cdot\vec{\nabla}_1 \vec{\sigma}_2\cdot\vec{\nabla}_1 \Delta^{\rm
  S}_{12} \big) \; + \;  {\cal O}\bigg(g^4, \, \frac{g^2}{m} \bigg) \,. \label{S:Minkowski:Space}
\eeq
With the action truncated at this level, the generating functional 
\beq
Z [\eta^\dagger,  \eta] \; = \; A \int DN^\dagger DN \, \exp \left(i S_N + i
   [\eta^\dagger_1 N_1 + N^\dagger_1 \eta_1] \right)
\label{GenFIncompl}
\eeq
gives rise to iterative, ladder-type $n$-nucleon diagrams with the
regularized static
one-pion exchange potential, which can be read off from the
instantaneous action $S_N$ in Eq.~(\ref{S:Minkowski:Space}) by taking functional derivatives with respect
to the nucleon fields.

\section{Loop contributions and the functional determinant}
\def\theequation{\arabic{section}.\arabic{equation}}
\label{sec:Determinant}

In the previous section, we have shown that our method 
successfully reproduces the lowest-order contribution to the 2N
potential generated by the one-pion exchange,  shown in
Fig.~\ref{fig2} (b). However, at the same
order $g^2$, there is also a contribution to the nucleon self-energy
as depicted in Fig.~\ref{fig2} (a). Per construction, the generating
functional in Eq.~(\ref{GenFIncompl}) cannot produce nonvanishing
contributions of such kind, signalling that the results obtained so
far are incomplete. As will be shown below, the still missing
contribution to the generating functional emerges from the functional
determinant generated by the employed field redefinitions. If the
considered transformations of the nucleon fields were local, the
determinant would be equal to one (at least, within $\zeta$-function
regularization).
However, for nonlocal transformation
this does not hold anymore, and one needs to calculate the functional
determinant explicitly.

Functional determinants play an important role in the path-integral
formalism and are discussed in many textbooks on quantum field
theory. For the sake of completeness, we derive in appendix
\ref{functional:det} the formal expression for the functional determinant
arising from a general redefinition of Grassmann fields in the path integral. 

We start with the generating functional $Z[\eta^\dagger, \eta]$ in
Eq.~(\ref{path:int:eucl}), written in terms of the Euclidean action $S_N^{\rm E}$
that involves non-instantaneous interactions. After performing nucleon
field redefinitions we obtain
\begin{multline}
\int D N D N^\dagger \exp\left(-S_N^{\rm E}+\int
  d^4x \,\left[\eta^\dagger N+N^\dagger\eta\right]\right)\\ = \int D
N^\prime D N^{\prime\dagger} \, {\rm Det}\left[\begin{array}{cc}
\frac{\delta N^\prime}{\delta N} & \frac{\delta N^\prime}{\delta
                                   N^\dagger} \\
\frac{\delta N^{\prime\dagger}}{\delta N} & \frac{\delta N^{\prime\dagger}}{\delta
                                   N^\dagger}
\end{array}\right]\exp\left(-S_{N(N^\prime)}^{\rm E}+\int
  d^4x \,\left[\eta^\dagger N(N^\prime,N^{\prime\dagger})+N^{\dagger}(N^\prime, N^{\prime\dagger})\eta\right]\right) \\ \simeq \int D
N^\prime D N^{\prime\dagger} \, {\rm Det}\left[\begin{array}{cc}
\frac{\delta N^\prime}{\delta N} & \frac{\delta N^\prime}{\delta
                                   N^\dagger} \\
\frac{\delta N^{\prime\dagger}}{\delta N} & \frac{\delta N^{\prime\dagger}}{\delta
                                   N^\dagger}
\end{array}\right]\exp\left(-S_{N(N^\prime)}^{\rm E}+\int
  d^4x \,\left[\eta^\dagger N^\prime+N^{\prime\dagger}\eta\right]\right),\label{funct:det:path:int}
\end{multline}
where $\simeq$ signifies that we neglect all induced terms involving
external sources like the ones given in
Eq.~(\ref{eta:higher:N}), which do not affect S-matrix elements. Further,
Eq.~(\ref{funct:det:path:int}) is only valid if all field
transformations are of the form $N (x) =N^\prime (x) + O_\alpha N^\prime (x) f_\beta
  [N^{\prime\dagger} O_\gamma N^\prime ]$ and
$N^\dagger(x) =N^{\prime\dagger} (x)+  N^{\prime\dagger} (x) O^\dagger_\alpha
g_\beta [N^{\prime\dagger}  O_\gamma N^\prime ]$,
where $O$ are some operators acting on the nucleon fields and
the functionals $f$ and $g$ are at least linear in the local field
products $N^{\prime\dagger} O_\gamma
   N^{\prime}$. 
For the case at hand, the
transformations have the form
\beqa
N_1&=&N_1'\; -\; \frac{g^2}{8 F^2}\, R^{\Lambda_\tau}_{13} N_3'
\, \fet{\tau}_3\cdot\fet{\tau}_2 \, \big(\vec{\sigma}_3\cdot\vec{\nabla}_3 \vec{\sigma}_2\cdot\vec{\nabla}_3 \partial_3^0\Delta^{\rm
  ES}_{32} \big),\nn
N^\dagger_1&=&N'^\dagger_1\; +\; \frac{g^2}{8 F^2}\, 
\, \big(\vec{\sigma}_3\cdot\vec{\nabla}_3 \vec{\sigma}_2\cdot\vec{\nabla}_3 \partial_3^0\Delta^{\rm
  ES}_{32} \big)\, \fet{\tau}_3\cdot\fet{\tau}_2 \, N'^\dagger_3  R^{\Lambda_\tau}_{31}.\label{fieldredef:compact}
\eeqa
Inverting the above equations, we obtain up to terms of the order
$g^2$:
\beqa
N_1'&=&N_1 \; +\; \frac{g^2}{8 F^2}\, R^{\Lambda_\tau}_{13} N_3
\, \fet{\tau}_3\cdot\fet{\tau}_2 \, \big(\vec{\sigma}_3\cdot\vec{\nabla}_3 \vec{\sigma}_2\cdot\vec{\nabla}_3 \partial_3^0\Delta^{\rm
  ES}_{32} \big),\nn
N'^\dagger_1&=&N^\dagger_1\; -\; \frac{g^2}{8 F^2}\, 
\, \big(\vec{\sigma}_3\cdot
\vec{\nabla}_3 \vec{\sigma}_2\cdot\vec{\nabla}_3  \partial_3^0\Delta^{\rm
  ES}_{32} \big)\, \fet{\tau}_3\cdot\fet{\tau}_2 \, N^\dagger_3  R^{\Lambda_\tau}_{31}.\label{fieldredef:compactNPrime}
\eeqa
To compute the functional determinant, we use the standard expression
\beqa
{\rm Det}[M]&=&\exp\big({\rm Tr}\log[M]\big)\,,
\eeqa
with the Jacobi matrix 
\beqa
M_{j k}&=&\left(\begin{array}{cc}
\frac{\delta N^\prime_j}{\delta N_k} & \frac{\delta N^\prime_j}{\delta
                                   N^\dagger_k} \\
\frac{\delta N^{\prime\dagger}_j}{\delta N_k} & \frac{\delta N^{\prime\dagger}_j}{\delta
                                   N^\dagger_k}
\end{array}\right).
\eeqa
To calculate the elements of $M$ we first rewrite Eqs.~(\ref{fieldredef:compactNPrime}) in a
more explicit form
\beqa
N^\prime_j&=&N_j+ \sumint\frac{g^2}{8
  F^2}[\sigma^a\fet{\tau} R^{\Lambda_\tau}]_{j
  n}\cdot[\sigma^b\fet{\tau}]_{r
  s}\, \big(\nabla_n^a \nabla_n^b \partial_n^0\Delta^{\rm ES}_{nr} \big) N_n N^\dagger_r N_s\,,\nn
N^{\prime\dagger}_j&=&N^{\prime\dagger}_j-\sumint\frac{g^2}{8
  F^2}[\sigma^a \fet{\tau} R^{\Lambda_\tau}]_{
  n j}\cdot[\sigma^b\fet{\tau}]_{s
  r} \big( \nabla_n^a \nabla_n^b \partial_n^0\Delta^{\rm ES}_{nr} \big) N^\dagger_n N^\dagger_s N_r\,,
\eeqa
where $a$ and $b$ refer to the corresponding Cartesian components and 
the subscripts $n$, $r$ and $s$ comprise the position, spin and
isospin indices of the nucleon fields. In the above expressions, the
integrations and summations are performed over the positions and
discrete quantum numbers of the nucleons, respectively.
The diagonal elements of $M$ are then easily obtained to be
\beqa
\frac{\delta N^\prime_j}{\delta N_k}&=&\delta_{j k} +
\sumint\frac{g^2}{8 F^2}[\sigma^a \fet{\tau}R^{\Lambda_\tau}]_{j
  n}\cdot[\sigma^b \fet{\tau}]_{r
  s}\, \big(\nabla_n^a \nabla_n^b \partial_n^0\Delta^{\rm ES}_{nr}\big) \left( N^\dagger_r
  N_s\delta_{n k}+N_n N^\dagger_r \delta_{s k}\right),\nn
\frac{\delta N^{\prime\dagger}_j}{\delta N^\dagger_k}&=&\delta_{j k}
-\sumint \frac{g^2}{8 F^2}[\sigma^a \fet{\tau}R^{\Lambda_\tau}]_{
  n j}\cdot[\sigma^b\fet{\tau}]_{s
  r} \, \big(\nabla_n^a \nabla_n^b  \partial_n^0\Delta^{\rm ES}_{nr} \big)\left(N^\dagger_s N_r\delta_{n
  k}-N^\dagger_n N_r\delta_{s k}\right).
\eeqa
To calculate the determinant, we use $\zeta$-function regularization~\cite{Gasser:1983yg,Ecker:1994pi,Meissner:1998rw}, see Eq.~(\ref{ZetaAFunctionMomentumSpace}), and
start with Fourier-transforming $M$ to momentum space:
\beqa
[M_{1,1}]_{p^\prime\alpha^\prime , \, 
  p\alpha}&=&\int d^4 x_j d^4 x_k \, e^{-ip^\prime\cdot x_j} \, e^{ip\cdot
  x_k} \, \frac{\delta N_{\alpha^\prime}^\prime (x_j) }{\delta
  N_{\alpha} (x_k)}\nn
&=& (2\pi)^4\delta(p^\prime-p)\delta_{\alpha^\prime \alpha} \; -\; i\frac{g^2}{8
  F^2}\int d^4 x_j d^4 x_k d^4 x_n d^4 x_r d^4x_s\frac{d^4
  q^\prime}{(2\pi)^4}\frac{d^4
  q}{(2\pi)^4}\frac{d^4l^\prime}{(2\pi)^4}\frac{d^4l}{(2\pi)^4} \nn
&&{}\times e^{-ip^\prime\cdot x_j} \, e^{ip\cdot
  x_k} \, e^{il\cdot (x_n-x_r)} \, e^{il^\prime\cdot(x_j-x_n)}\, e^{-iq^\prime\cdot x_r}
\, \tilde\Delta^{\rm ES}_l \, \tilde{R}^{\Lambda_\tau}_{l^\prime} \, l_0
\, \delta(x_r-x_s)\nn
&&{}\times \bigg\{
[\vec{\sigma}\cdot\vec{l}\,\fet{\tau}]_{\alpha^\prime, \alpha}\cdot \tilde{N}^\dagger_{q^\prime}\vec{\sigma}\cdot\vec{l}\,\fet{\tau}\tilde
N_q \, e^{iq\cdot x_s}\, \delta(x_n-x_k) \; +\; 
[\vec{\sigma}\cdot\vec{l}\,\fet{\tau} \tilde{N}_q]_{\alpha^\prime}\cdot
[\tilde{N}^\dagger_{q^\prime}\vec{\sigma}\cdot\vec{l}\,\fet{\tau}]_{\alpha}
\, e^{iq\cdot x_n}\delta(x_s-x_k)\bigg\}\nn
&=& (2\pi)^4\delta(p^\prime-p)\delta_{\alpha^\prime \alpha}
\; -\; i\frac{g^2}{8
  F^2}\tilde{R}^{\Lambda_\tau}_{p^\prime}\int\frac{d^4 q^\prime}{(2\pi)^4}\frac{d^4
  q}{(2\pi)^4}(2\pi)^4\delta(-p^\prime-q^\prime+p+q)\nn
&&{}\times\bigg\{\tilde\Delta^{\rm ES}_{q^\prime-q} \,
(q_0-q_0^\prime) \, 
[\vec{\sigma}\cdot(\vec{q}-\vec{q}^{\,\prime})\,\fet{\tau}]_{\alpha^\prime,\alpha}\cdot \tilde{N}^\dagger_{q^\prime}\vec{\sigma}\cdot(\vec{q}-\vec{q}^{\,\prime})\,\fet{\tau}\tilde
N_q\nn
&&\hspace{0.7cm}{}+ \tilde\Delta^{\rm ES}_{q^\prime-p} \,
(p_0-q_0^\prime) \, 
[\vec{\sigma}\cdot(\vec{p}-\vec{q}^{\,\prime})\,\fet{\tau} \tilde{N}_q]_{\alpha^\prime}\cdot
[\tilde{N}^\dagger_{q^\prime}\vec{\sigma}\cdot(\vec{p}-\vec{q}^{\,\prime})\,\fet{\tau}]_{\alpha}\bigg\},
\eeqa
where $\alpha$ and $\alpha '$ collect the spin and isospin
indices of the nucleon field 
and
$\tilde f$ refers to the Fourier-transform of
a coordinate-space quantity $f$. In particular,
\beqa
\tilde{R}^{\Lambda_\tau}_p \; \equiv \; \tilde{R}^{\Lambda_\tau}(p) \; =\;
\int d^4 x \,  e^{-i p\cdot x} \, R^{\Lambda_\tau} (x) \;=\;\frac{\Lambda_\tau}{ip_0+\Lambda_\tau}.
\eeqa
In a completely analogous way, we obtain for the second diagonal
element of $M$: 
\beqa
[M_{2,2}]_{p^\prime \alpha^\prime, \, 
  p \alpha}&=&\int d^4 x_j d^4 x_k \, e^{-ip^\prime\cdot x_j} \, e^{ip\cdot
  x_k} \, \frac{\delta N_{\alpha^\prime}^{\prime\dagger} (x_j) }{\delta
  N_{\alpha}^\dagger (x_k) } \nn
&=& (2\pi)^4\delta(p^\prime-p)\delta_{\alpha^\prime \alpha}
\; + \; i\frac{g^2}{8
  F^2}\tilde{R}^{\Lambda_\tau}_{-p^\prime}\int\frac{d^4 q^\prime}{(2\pi)^4}\frac{d^4
  q}{(2\pi)^4}(2\pi)^4\delta(p^\prime+q^\prime-p-q)\nn
&&{}\times \bigg\{\tilde\Delta^{\rm ES}_{q^\prime-q} \,
(q_0-q_0^\prime) \,
[\vec{\sigma}\cdot(\vec{q}-\vec{q}^{\,\prime})\,\fet{\tau}]_{\alpha
  , \alpha^\prime}\cdot  \tilde{N}^\dagger_{q^\prime}\vec{\sigma}\cdot(\vec{q}-\vec{q}^{\,\prime})\,\fet{\tau}\tilde
N_q\nn
&&\hspace{0.6cm}{} + \tilde\Delta^{\rm ES}_{q+p} (q_0+p_0)
[\vec{\sigma}\cdot(\vec{q}+\vec{p} \, )\,\fet{\tau} \tilde{N}_q]_{\alpha}\cdot
[\tilde{N}^\dagger_{q^\prime}\vec{\sigma}\cdot(\vec{q}+\vec{p}\, )\,\fet{\tau}]_{\alpha^\prime}\bigg\}\,.
\eeqa
To use the $\zeta$-function regularization \cite{Reuter:1984kw}, we multiply the matrix $M$ by
$(p^\prime)^2+\epsilon^2$. That is, instead of calculating ${\rm Det}\big[M\big]$,
  we calculate ${\rm Det}\big[{\cal D} M\big]$ and ${\rm
    Det}\big[{\cal D}]$,
where
\beqa
[{\cal D}]_{x \alpha', y \alpha} &=&
(-\partial_x^2+\epsilon^2)\delta^4(x-y)\delta_{\alpha^\prime
  \alpha}\,.
\eeqa
The specific form of this operator is not important, it should only be an
elliptic operator which has a discrete spectrum.
The original determinant is then given by 
\beqa
{\rm Det}\big[M\big]&=&\frac{{\rm Det}\big[{\cal D} M\big]}{{\rm
    Det}\big[{\cal D}]}.
\eeqa
Evaluating the trace in the spin-isospin space we find
\beqa
&& \sum_{\alpha}\big([{\cal D}M_{1,1}]_{p \alpha , \, p \alpha}+[{\cal
  D}M_{2,2}]_{p \alpha , \, p \alpha}\big)\; =\; 
4 (2\pi)^4\delta(p^\prime -p ) (p^2+\epsilon^2) \Big|_{p^\prime \to p}
\; -\;i\frac{g^2}{8 F^2}
\int\frac{d^4
  q}{(2\pi)^4}(p^2+\epsilon^2)\\
&& \hspace{1.5cm}{} \times  \Big\{\tilde{R}^{\Lambda_\tau}_{-p}\, \tilde\Delta^{\rm
  ES}_{q+p} \, (q_0+p_0) \, 
\tilde{N}^\dagger_q[\vec{\sigma}\cdot(\vec{q}+\vec{p} \, )\,\fet{\tau}]^2
\tilde{N}_q \,-\, \tilde{R}^{\Lambda_\tau}_{p} \, \tilde\Delta^{\rm ES}_{q-p} \,  (p_0-q_0)
\, \tilde{N}^\dagger_q[\vec{\sigma}\cdot(\vec{p}-\vec{q} \, )\,\fet{\tau}]^2
\tilde{N}_q \Big\}.
\nonumber
\eeqa
Inserting this relation in Eq.~(\ref{ZetaAFunctionMomentumSpace}), we
obtain for the generalized $\zeta$-function (see Appendix
\ref{zetafunctionReg} for the various definitions)
\beqa
\zeta_{{\cal D}M}(\sigma)&=&\zeta_{{\cal D}M}^{(0)}(\sigma)+\zeta_{{\cal
    D}M}^{(2)}(\sigma),
\eeqa
where
\beqa
\zeta_{{\cal D}M}^{(0)}(\sigma)&=&\int \frac{d^4
  p}{(2\pi)^4}\frac{\mu^{2 \sigma}}{\big(p^2+\epsilon^2\big)^\sigma}4
V,\nn
\zeta_{{\cal D}M}^{(2)}(\sigma)&=&-\int \frac{d^4
  p}{(2\pi)^4}\frac{\sigma\mu^{2
    \sigma}}{\big(p^2+\epsilon^2\big)^{\sigma+1}}\sum_{\alpha}\big([{\cal
  D}M_{1,1}]_{p \alpha , \, p \alpha}+[{\cal
  D}M_{2,2}]_{p\alpha , \, p \alpha}\big),
\eeqa
with $V=\int d^4x$ denoting the Euclidean space-time volume and 
the superscript $n$ of $\zeta_{{\cal D}M}^{(n)}$ referring to the power of $g$.
Since $\zeta_{{\cal D}M}^{(0)}$ contributes only to the normalization of the
path integral, it needs not be considered any further. We thus
concentrate on $\zeta_{{\cal D}M}^{(2)}$, whose explicit form is given
by
\beqa
\zeta_{{\cal D}M}^{(2)}(\sigma)&=&3\frac{g^2}{4 F^2}\int \frac{d^4
  q}{(2\pi)^4} \tilde{N}^\dagger_q\tilde{N}_q \, i\int \frac{d^4
    p}{(2\pi)^4}\frac{\sigma\mu^{2 \sigma}}{\big((p-q)^2+\epsilon^2\big)^{\sigma}}\frac{\Lambda_\tau}{i(q_0-p_0)+\Lambda_\tau}
 \,  p_0 \, \vec{p}^{\,2}\, \tilde \Delta^{\rm ES}_p.
\eeqa
The expression for ${\rm Tr}\,\log({\cal D} M)$ can be obtained by
calculating the derivative of $\zeta_{{\cal
    D}M}^{(2)}(\sigma)$ and taking the limit $\sigma\to 0^+$, see
Eq.~(\ref{AppBtemp1}). Since the
integral is well behaved at $\sigma=0$, it is legitimate to interchange the integration
and the limit operation, and we obtain
\beqa
\zeta_{{\cal D}M}^{(2)\,\prime}(0^+)&=&\int \frac{d^4
  q}{(2\pi)^4} \, \tilde{N}^\dagger_q\tilde{N}_q \, \tilde{I}_{\rm ES}(q_0),
\eeqa
where the loop integral $\tilde{I}_{\rm ES}(q_0)$ is given by
\beqa
\tilde{I}_{\rm ES}(q_0)&=&3\frac{g^2}{4 F^2}i\int \frac{d^4
    p}{(2\pi)^4}\frac{\Lambda_\tau}{i(q_0-p_0)+\Lambda_\tau}
  \, p_0 \, \vec{p}^{\,2}\, \tilde \Delta^{\rm ES}_p.\label{IES:SymbolicForm}
\eeqa
To calculate the integral we insert in
Eq.~(\ref{IES:SymbolicForm}) the explicit form of the propagator
\beqa
\tilde{\Delta}^{\rm
  ES}_p&=&-\frac{1}{p_0^2}\Bigg(\frac{e^{-\frac{p_0^2+\vec p \, ^2+M^2}{\Lambda^2}}}{p_0^2+\vec p \, ^2+M^2}-\frac{e^{-\frac{\vec p \, ^2+M^2}{\Lambda^2}}}{\vec p \, ^2+M^2}\Bigg). 
\eeqa
In addition, as already advertised in the previous section, we remove
the temporal regulator by taking the limit $\Lambda_\tau\to\infty$
and obtain  
\beqa
\tilde{I}_{\rm ES}(q_0)&=&3\frac{g^2}{4
  F^2}\lim_{\Lambda_\tau\to\infty}\int\frac{d^4
  p}{(2\pi)^4}\frac{i\Lambda_\tau  \vec p \, ^2 }{i(p_0+q_0)+\Lambda_\tau}\frac{1}{p_0}\Bigg(\frac{e^{-\frac{p_0^2+\vec p \, ^2+M^2}{\Lambda^2}}}{p_0^2+\vec p \, ^2+M^2}-\frac{e^{-\frac{\vec p \, ^2+M^2}{\Lambda^2}}}{\vec p \, ^2+M^2}\Bigg).
\eeqa
We rewrite this expression as
\beqa
\tilde{I}_{\rm ES}(q_0)&=&3\frac{g^2}{4
  F^2}\lim_{\Lambda_\tau\to\infty}\int\frac{d^4
  p}{(2\pi)^4}\frac{i\Lambda_\tau \vec p \, ^2
}{i(p_0+q_0)+\Lambda_\tau}\frac{1}{p_0}\Bigg(\frac{e^{-\frac{p_0^2+\vec p \, ^2+M^2}{\Lambda^2}}}{p_0^2+\vec p \, ^2+M^2}-\frac{e^{-\frac{\vec p \, ^2+M^2}{\Lambda^2}}}{p_0^2+\vec p \, ^2+M^2}\Bigg)\nn
&& +\;3\frac{g^2}{4
  F^2}\lim_{\Lambda_\tau\to\infty}\int\frac{d^4
  p}{(2\pi)^4}\frac{i\Lambda_\tau \vec p \, ^2
}{i(p_0+q_0)+\Lambda_\tau}\frac{1}{p_0}\Bigg(\frac{e^{-\frac{\vec p \, ^2+M^2}{\Lambda^2}}}{p_0^2+\vec p \, ^2+M^2}-\frac{e^{-\frac{\vec p \, ^2+M^2}{\Lambda^2}}}{\vec p \, ^2+M^2}\Bigg).\label{IESLimit}
\eeqa
The integral in the first line of Eq.~(\ref{IESLimit})
is well-defined even if one takes the limit $\Lambda_\tau\to\infty$ for
the integrand. Since the integrand is an odd function
in $p_0$ in the limit $\Lambda_\tau\to\infty$, the first line of
Eq.~(\ref{IESLimit}) yields vanishing result. The integral in the second line of Eq.~(\ref{IESLimit})
can be calculated analytically:
\beqa
\tilde{I}_{\rm ES}(q_0)&=&-\frac{3 g^2}{8
  F^2}\lim_{\Lambda_\tau\to\infty}\int \frac{d^3 p}{(2\pi)^3}\frac{e^{-\frac{\omega_p^2}{\Lambda^2}}}{\omega_p^2}\frac{\vec p \, ^2 \,\Lambda_\tau}{iq_0+\Lambda_\tau+\omega_p}\;=\; -\frac{3 g^2}{8
  F^2}\int \frac{d^3
  p}{(2\pi)^3}\, \vec p \, ^2\, \frac{e^{-\frac{\omega_p^2}{\Lambda^2}}}{\omega_p^2}\nn
&=&-\frac{3 g^2}{64 F^2 \pi^2}\left[2\pi M^3 {\rm
    erfc}\left(\frac{M}{\Lambda}\right)+e^{-\frac{M^2}{\Lambda^2}}\sqrt{\pi}\Lambda\left(\Lambda^2-2
    M^2\right)\right]\,,
\label{SelfEnergyI}
\eeqa
where $\omega_p=\sqrt{\vec p^{\,2}+M^2}$ and the complementary error function is defined by
\beqa
{\rm erfc} (z)&=&1-{\rm erf}(z), \quad {\rm
  erf}(z)\,=\,\frac{2}{\sqrt{\pi}}\int_0^z dt\, e^{-t^2}.
\eeqa
Notice that the calculated loop integral $\tilde{I}_{\rm ES}$
does not depend on $q_0$. Now, changing to the short-hand notation
introduced in the previous section, our final result for the
functional determinant takes the form
\beqa
\lim_{\Lambda_\tau\to\infty}{\rm Tr}\log\left({\cal D}M\right)
&=&-\lim_{\Lambda_\tau\to\infty}\zeta_{{\cal D}M}^\prime(0^+)\,=\,{\rm
  const}-N^{\dagger}_1N_1\tilde{I}_{\rm ES},\label{selfenergycontr:reg:limit}
\eeqa
so that the corrected expression for the instantaneous action in the limit $\Lambda_\tau\to\infty$ is
given by
\beqa
S_N^{\rm E}&=&N^\dagger_1\bigg(\partial_1^0 -\frac{\vec{\nabla}_1^{\,2}}{2
    m}+\tilde{I}_{\rm ES}\bigg)N_1 \; + \; \frac{g^2}{8 F^2}\fet{\tau}_1\cdot\fet{\tau}_2
\big( \sigma_1 \cdot \nabla_1 \,  \sigma_2\cdot \nabla_1
\Delta^{\rm
  S}_{12} \big) \; + \;  {\cal O}\bigg(g^4, \, \frac{g^2}{m} \bigg) .\label{euclactiong2Instant}
\eeqa
After performing the Wick rotation, we finally obtain the action in Minkowski space:
\beqa
S_N&=&N^\dagger_1\bigg(i\partial_1^0 +\frac{\vec{\nabla}_1^{\,2}}{2
    m}-\tilde{I}_{\rm ES}\bigg)N_1 \; - \; \frac{g^2}{8 F^2}\fet{\tau}_1\cdot\fet{\tau}_2
\big( \sigma_1 \cdot \nabla_1\,  \sigma_2 \cdot \nabla_1
\Delta^{\rm
  S}_{12} \big) \; + \;  {\cal O}\bigg(g^4, \, \frac{g^2}{m} \bigg).\label{euclactiong2Minkowski}
\eeqa
Clearly, the quantity $\tilde{I}_{\rm ES}$ is nothing but the
nucleon self-energy in the (regularized) theory with redefined
nucleon fields, specified by the generating functional in Eq.~(\ref{GenFIncompl})
with the instantaneous action $S_N$ given in Eq.~(\ref{euclactiong2Minkowski}). Since $\tilde{I}_{\rm ES}$
is just a constant, it only gives rise to a shift of the nucleon
mass. Expanding $\tilde{I}_{\rm ES}$ in inverse powers
of the cutoff $\Lambda$, we obtain  
\beqa
\tilde{I}_{\rm ES}&=&-\frac{3 g^2}{64
  \pi^{3/2} F^2}\Lambda^3+\frac{9 g^2 M^2}{64 \pi^{3/2} F^2}\Lambda -
\frac{3 g^2 M^3}{32 \pi F^2}+{\cal O}\bigg(\frac{1}{\Lambda } \bigg),\label{nuclmassrenorm}
\eeqa
where the first two terms analytic in $M^2$ are scheme dependent. The last 
term $\propto M^3$, on the other hand, reproduces the well-known leading
non-analytic contribution to the nucleon mass \cite{Gasser:1979hf},
which is unambiguously defined and model-independent.

\section{The nucleon $Z$-factor}
\def\theequation{\arabic{section}.\arabic{equation}}
\label{sec:ZFactor}

In the last two sections, we have worked out the instantaneous action
to order $g^2$, which correctly reproduces the leading non-analytic
contribution to the nucleon mass shift and the one-pion exchange
two-nucleon potential. Since $\tilde{I}_{\rm ES}$ is independent of
$q_0$, see Eq.~(\ref{SelfEnergyI}), the generating functional in Eq.~(\ref{GenFIncompl}) with the
instantaneous action $S_N$ given in Eq.~(\ref{euclactiong2Minkowski}) cannot produce any
non-vanishing shift of the nucleon $Z$-factor. On the other
hand, the path integral we have started from in
Eqs.~(\ref{Generating:Functional:Eucl}) and (\ref{Regularized:LE}) or,
equivalently, in Eqs.~(\ref{path:int:eucl}) and (\ref{NN:NonInstant})
after integrating over the pion fields does lead to a non-vanishing
contribution to the nucleon $Z$-factor at order $g^2$. This mismatch
does, of course, not indicate any inconsistency given that
$Z$-factors, in contrast to S-matrix elements, 
depend on the choice of interpolating fields and are not
observable. The different result for the nucleon $Z$-factor originates
from our choice to ignore source-dependent terms induced 
by the nucleon field redefinitions, see
Eq.~(\ref{eta:higher:N}). Indeed, had we taken into account the terms
in Eq.~(\ref{eta:higher:N}), the resulting generating functional would simply be identical to the
original one without nucleon field redefinitions, thus producing the
same $Z$-factor (and Green's functions). Below, we demonstrate by
explicit 
calculations that taking into account the terms in
Eq.~(\ref{eta:higher:N})
indeed allows one to reproduce the nucleon $Z$-factor at
the one-loop accuracy level. This provides another nontrivial consistency
check of the approach. 

We focus on the contributions to the nucleon two-point function
induced by the neglected terms in Eq.~(\ref{eta:higher:N}), as
signified with the superscript ``ind'': 
\beqa
G^{\rm ind}(x_1, x_4) &=& G^{\rm ind}(x_1 - x_4) \; = \;  \frac{1}{Z[0]} \, \frac{\delta^2}{\delta
  \eta (x_4)  \, \delta \eta^\dagger (x_1)} Z^{\rm ind}[\eta^\dagger,
\eta] \bigg|_{\eta^\dagger = \eta = 0} \nn
&=&-3\frac{g^2}{8 F^2} \sumintnum \bigg\{
[R^{\Lambda_\tau}]_{12}
\wick{1}{<1N_2>1N^\dagger_3}\wick{1}{
 <1N_3
 >1N^\dagger_4} \, \big(\vec{\nabla}_2\cdot\vec{\nabla}_2 \, \partial_2^0\Delta^{\rm
 ES}_{23} \big) \;  -\;  
\wick{1}{<1N_1>1N^\dagger_2}\wick{1}{<1N_2>1N^\dagger_3}[R^{\Lambda_\tau}]_{34}\, 
\big(\vec{\nabla}_3\cdot\vec{\nabla}_3\, \partial_3^0\Delta^{\rm
  ES}_{32} \big)\bigg\}\,, \nonumber
\eeqa
where only terms of the order $\mathcal{O} (g^2 )$ are kept. Here, we use the familiar notation for the Wick-contractions
\beqa
\wick{1}{<1N_1>1N^\dagger_2} &=& \lim_{m\to\infty}P^{\rm E}(x_1-x_2)\,,
\eeqa
where $P^{\rm E}$ is a free Euclidean nucleon propagator defined in Eq.~(\ref{FreeEuclideanNucleonPropagator}).
To read out the induced contribution to the self-energy integral
$\tilde{I}_{\rm ES}^{\rm ind}(p)$, we Fourier-transform $G^{\rm
  ind}(x_1 - x_4)$ to momentum space and obtain\footnote{Since the
  order-$g^2$ contribution to the two-point function is one-particle
  irreducible, it directly corresponds to the nucleon self energy.}
\beqa
-(2\pi)^4\delta(p^\prime-p) \, \frac{\tilde{I}_{\rm ES}^{\rm
    ind}(p)}{(ip_0+\epsilon)^2}
& =&\int d^4 x_1 d^4
x_4 \, e^{-ip^\prime\cdot x_1+ip\cdot x_4}
\, G^{\rm  ind}(x_1 - x_4) \nn
&=& 3\frac{g^2}{8 F^2}\int \prod_{j=1}^3 \prod_{k=1}^4 d^4 x_k \frac{d^4
  p_j}{(2\pi)^4} \frac{d^4 l}{(2\pi)^4}\, 
e^{-ip^\prime\cdot x_1+ip\cdot x_4}\nn
&& {} \times  \bigg[e^{ip_1\cdot (x_1-x_2)}e^{ip_2\cdot
  (x_2-x_3)}e^{ip_3\cdot (x_3-x_4)}e^{il\cdot
  (x_2-x_3)}\, \prod_{k=1}^3
\prod_{j=2}^3\, \frac{1}{ip_j^0+\epsilon}\frac{\Lambda_\tau}{ip_k^0+\Lambda_\tau}
\, i  \vec l^{\; 2}
 l_0 \, \tilde{\Delta}^{\rm ES}(l)\nn
&& {} -  e^{ip_1\cdot (x_1-x_2)}e^{ip_2\cdot
  (x_2-x_3)}e^{ip_3\cdot (x_3-x_4)}e^{il\cdot
  (x_3-x_2)}\, \prod_{j=1}^3 \prod_{k=1}^2\,
\frac{1}{ip_k^0+\epsilon}\frac{\Lambda_\tau}{ip_j^0+\Lambda_\tau}\, i \vec l^{\; 2} 
l_0\, \tilde{\Delta}^{\rm ES}(l)\bigg]
\nn
&=&-3\frac{g^2}{8
  F^2}(2\pi)^4\delta(p^\prime-p)\,
\frac{1}{ip_0+\epsilon}\left(\frac{\Lambda_\tau}{ip_0+\Lambda_\tau}\right)^2
\int
\frac{d^4
  l}{(2\pi)^4}
\frac{i \vec l^{\; 2}}{l_0}\Bigg(\frac{e^{-\frac{l_0^2+\vec l^{\, 2}+M^2}{\Lambda^2}}}{l_0^2+\vec l^{\, 2}+M^2}-\frac{e^{-\frac{\vec l^{\, 2}+M^2}{\Lambda^2}}}{\vec l^{\, 2}+M^2}\Bigg)
\nn
&& {} \times \bigg[\frac{1}{i(p_0-l_0)+\epsilon}\, \frac{\Lambda_\tau}{i(p_0-l_0)+\Lambda_\tau}
\; - \;
\frac{1}{i(p_0+l_0)+\epsilon}\, \frac{\Lambda_\tau}{i(p_0+l_0)+\Lambda_\tau}\bigg].\label{ZNFactorDerivation}
\eeqa
Note that the sign on the left hand side of Eq.~(\ref{ZNFactorDerivation})
comes from the Euclidean version of the self-energy. In Euclidean space, the
full nucleon propagator is given by $1/(i p_0 + \Sigma_{\rm E}(p))$, where
$\Sigma_{\rm E}(p)$ is the nucleon self-energy.
From Eq.~(\ref{ZNFactorDerivation}) we see that $\tilde{I}_{\rm ES}^{\rm
    ind}(p)$
can be calculated without the temporal regulator.  We, therefore, take
the limit
$\Lambda_\tau\to\infty$ in the integrand and obtain
\beqa
\lim_{\Lambda_\tau\to\infty} \tilde{I}_{\rm ES}^{\rm
    ind}(p) &=&-3\frac{g^2}{4
  F^2}ip_0\int \frac{d^4
  l}{(2\pi)^4}\frac{\vec l^{\; 2}}{l_0^2-(p_0-i\epsilon)^2}\Bigg(\frac{e^{-\frac{l_0^2+\vec l^{\, 2} +M^2}{\Lambda^2}}}{l_0^2+\vec l^{\, 2}+M^2}-\frac{e^{-\frac{\vec l^{\, 2}+M^2}{\Lambda^2}}}{\vec l^{\, 2}+M^2}\Bigg).\label{deltaTildeIES}
\eeqa
Notice that since the integral in the above equation  is well-defined
for $p_0=0$, we have $\lim_{\Lambda_\tau\to\infty} \tilde{I}_{\rm ES}^{\rm
    ind}(0) = 0$. Therefore, as expected, the induced contribution to the
  self-energy does not affect the nucleon mass, but
  only produces a shift of the nucleon $Z$-factor.

We still need to verify that the obtained one-loop correction to the
$Z$-factor agrees with the one obtained from the original,
non-instantaneous action. To this aim we compute the one-loop contribution to the two-point
function
generated by the action in Eq.~(\ref{NN:NonInstant}):
\beqa
G (x_1-x_4)&=&-\frac{g^2}{8 F^2} \sumintnum
\fet{\tau}_2\cdot\fet{\tau}_3 \big( \sigma_2 \cdot \nabla_2
\sigma_3 \cdot \nabla_2 \Delta^{\rm E}_{23} \big)\bigg(
\wick{111}{<1N_1>1N^\dagger_2  <2N_2
  >2N^\dagger_3<3N_3>3N^\dagger_4}
+\wick{123}{<1 N_1 <2N^\dagger_2 <3N_2
  >1N^\dagger_3>2N_3>3N^\dagger_4}\bigg)\nn
&=&-3\frac{g^2}{4 F^2}\sumintnum
\wick{111}{<1N_1>1N^\dagger_2 <2N_2
  >2N^\dagger_3<3N_3>3N^\dagger_4} \big(
\vec{\nabla}_2\cdot\vec{\nabla}_2\Delta^{\rm E}_{23} \big)\,.
\eeqa
To obtain the self-energy $\Sigma_{\rm E} (p)$, we Fourier-transform
the two-point function to momentum space. For the case of the
non-instantaneous action, there is no need to introduce a temporal
regulator. We, therefore, work from the beginning in the limit
$\Lambda_\tau \to \infty$ and obtain:
\beqa
-(2\pi)^4\delta(p^\prime-p)\, \frac{ \Sigma_{\rm E} (p)
}{(ip_0+\epsilon )^2}&=&\int d^4 x_1 d^4 x_4
e^{-ip^\prime\cdot x_1+ip\cdot x_4} G (x_1-x_4)\nn
&=&3\frac{g^2}{4 F^2}\int
\prod_{j=1}^3 \prod_{k=1}^4 d^4 x_k \frac{d^4 p_j}{(2\pi)^4} 
\frac{d^4 l}{(2\pi)^4}
e^{-ip^\prime\cdot x_1+ip\cdot x_4}
\nn
&& {} \times
e^{ip_1\cdot(x_1-x_2)}e^{ip_2\cdot(x_2-x_3)}e^{ip_3\cdot(x_3-x_4)}e^{i
l\cdot(x_2-x_3)}\prod_{j=1}^3\frac{1}{ip_j^0+\epsilon}
\vec l^{\; 2} \tilde{\Delta}^{\rm E} (l) \nn
&=&(2\pi)^4\delta(p^\prime-p)\left(\frac{1}{ip_0+\epsilon}\right)^2 3\frac{g^2}{4 F^2}\int\frac{d^4l}{(2\pi)^4}\frac{\vec l^{\, 2}}{i(p_0-l_0)+\epsilon}\frac{e^{-\frac{l_0^2+\vec l^{\, 2}+M^2}{\Lambda^2}}}{l_0^2+\vec l^{\, 2}+M^2}.
\eeqa
Expanding the self-energy contribution $\Sigma_{\rm E} (p)$ around
$p_0=0$, we finally obtain
\beqa
\Sigma_{\rm E} (p)&=&\Sigma_{\rm E} (0)
\;   -\; 3 ip_0\frac{g^2}{4 F^2} \int \frac{d^4
    l}{(2\pi)^4}\frac{\vec l^{\, 2}}{(l_0-i\epsilon)^2}\frac{e^{-\frac{l_0^2+\vec l^{\, 2}+M^2}{\Lambda^2}}}{l_0^2+\vec l^{\, 2}+M^2}\nn
&=&\Sigma_{\rm E} (0)\; -\; 3 ip_0\frac{g^2}{4
  F^2} 
\int \frac{d^4
    l}{(2\pi)^4}\frac{\vec l^{\, 2}}{(l_0-i\epsilon)^2}\Bigg(\frac{e^{-\frac{l_0^2+\vec l^{\, 2}+M^2}{\Lambda^2}}}{l_0^2+\vec l^{\, 2}+M^2}-\frac{e^{-\frac{\vec l^{\, 2}+M^2}{\Lambda^2}}}{\vec l^{\, 2}+M^2}\Bigg)\nn
&=&\Sigma_{\rm E} (0)\; +\; \lim_{\Lambda_\tau\to\infty}\tilde{I}_{\rm
  ES}^{\rm ind}(p)+{\cal O}((p^0)^2).\label{ZNFaktorReproduced}
\eeqa
This demonstrates that the expression for the $Z$-factor in the
original formulation is indeed restored when  keeping
the neglected contributions in Eq.~(\ref{eta:higher:N}). Notice
further that
\beqa
\Sigma_{\rm E}(0)&=&-3\frac{g^2}{4
  F^2}\int\frac{d^4l}{(2\pi)^4}\, \frac{\vec l^{\,
    2}}{il_0+\epsilon}\, \frac{e^{-\frac{l_0^{2}+\vec l^{\,
        2}+M^2}{\Lambda^2}}}{l_0^2+\vec l^{\, 2}+M^2}\nn
&=&3\frac{g^2}{4 F^2}\int \frac{d^4
  l}{(2\pi)^4} \, \frac{i\,\vec{l}^{\, 2}}{l_0-i\,\epsilon}\,
\frac{1}{l_0^2+\vec{l}^{\, 2}+M^2}\bigg(e^{-\frac{l_0^2+\vec{l}^{\,
      2}+M^2}{\Lambda^2}}-e^{-\frac{\vec{l}^{\, 2}+M^2}{\Lambda^2}}\bigg)\nn
&+&3\frac{g^2}{4 F^2}\int \frac{d^4
  l}{(2\pi)^4}\, \frac{i\,\vec{l}^{\, 2}}{l_0-i\,\epsilon}\,
\frac{e^{-\frac{\vec{l}^{\, 2}+M^2}{\Lambda^2}}}{l_0^2+\vec{l}^{\, 2}+M^2}\,.\label{SigmaEuclideanAtZeroPrepare}
\eeqa
The first term on the rhs of Eq.~(\ref{SigmaEuclideanAtZeroPrepare})
vanishes in the $\epsilon=0$ limit since the integrand has no singularity
at $l_0=0$ and is antisymmetric in $l_0$. The second term can be
calculated with the help of residue theorem:
\beqa
\Sigma_{\rm E}(0)&=&3\frac{g^2}{4 F^2}\int \frac{d^4
  l}{(2\pi)^4}\frac{i\,\vec{l}^{\,
    2}}{l_0-i\,\epsilon}\frac{e^{-\frac{\vec{l}^{\,
        2}+M^2}{\Lambda^2}}}{l_0^2+\vec{l}^{\, 2}+M^2}\;=\;-3\frac{g^2}{8 F^2}\int \frac{d^3
  l}{(2\pi)^3}\frac{\vec{l}^{\, 2}
  e^{-\frac{\vec{l}^{\, 2}+M^2}{\Lambda^2}}}{\vec{l}^{\, 2}+M^2}\;=\;{\tilde
  I}_{{\rm ES}}\,.
\eeqa
This  shows
that the nucleon mass shift, whose non-analytic part represents an
observable quantity,
coinsides in both formulations regardless
of whether or not the induced terms in Eq.~(\ref{eta:higher:N}) are
kept. 

The above considerations suggest an alternative (but equivalent)
formulation of our approach. Instead of performing nonlocal
redefinitions of the nucleon fields, we can make use of the freedom in
the choice of the external sources $\eta$ and
$\eta^\dagger$. Specifically, starting from the generating functional
in Eq.~(\ref{path:int:eucl}) after performing the integration
over the pion field, we vary the sources by replacing
\beq
\eta (x) \; \to \; \eta (x) + O_\alpha \eta (x) F_\beta [N^\dagger
O_\gamma N]\,, \quad \quad
\eta^\dagger (x) \; \to \; \eta^\dagger (x) + \eta^\dagger (x) O_\alpha G_\beta [N^\dagger
O_\gamma N]\,,
\eeq
where $O$ are some operators and the yet-to-be-determined functionals $F$ and $G$ are at least linear in the local
field products $N^\dagger O_\gamma N$. As already pointed out above, such modifications do not
affect S-matrix elements. Next, the induced terms in the generating
functional that are 
nonlinear in the fields are eliminated
via suitable field redefinitions 
$N (x) \; \to \; N (x) - O_\alpha N (x) G_\beta [N^\dagger
O_\gamma N] + \ldots$ and $N^\dagger (x) \; \to \; N^\dagger (x) - N^\dagger (x) O_\alpha F_\beta [N^\dagger
O_\gamma N] + \ldots$. This puts the modified generating
functional back to the
standard form with the external sources coupling linearly to the fields.
Finally, the functionals $F$ and $G$ are varied
until the interaction part of the action acquires an instantaneous
form.

\section{Summary and conclusions}
\def\theequation{\arabic{section}.\arabic{equation}}
\label{sec:summary}

In this paper we have introduced a new method for the derivation of
nuclear forces and current operators from the effective chiral
Lagrangian. Our study represents a first step towards deriving 
consistently regularized nuclear interactions in chiral EFT, which 
requires the cutoff to be introduced already at the
level of the effective Lagrangian. Because of the desired manifestly
Lorentz-invariant treatment of pion dynamics in chiral EFT, the regularized
Lagrangian involves arbitrarily high-order time derivatives of the pion
fields. This complicates the canonical quantization of the theory and
prevents one from using the established techniques such as TOPT or the method of unitary
transformation to derive nuclear potentials. The natural way around
this problem is to employ path-integral quantization. While the path integral
approach is the standard calculational tool in quantum field theory,
it has, to the best of our knowledge, not yet been used
for the
derivation of nuclear potentials. Here, we fill this gap and show how to employ the
path integral approach to derive nuclear forces in chiral EFT.

To keep the presentation simple, we have restricted ourselves in this
paper to a Yukawa-type model for non-relativistic nucleons interacting
with pions via a pseudovector coupling. This theory may serve
as a toy model of chiral EFT, and it does correctly reproduce the
one-pion exchange two-nucleon potential. After performing the Wick
rotation to Euclidean space-time, the pion kinetic term in the
Lagrangian was modified by introducing a Gaussian-type regulator,
whose form was chosen to match the regulator used in
Ref.~\cite{Reinert:2017usi}. The corresponding cutoff $\Lambda$ is
kept finite throughout the calculation. We also introduced a
temporal regulator in the nucleon kinetic term to be able to evaluate
loop integrals over zeroth momentum components. The temporal
cutoff $\Lambda_\tau$ is removed at the end of the calculation by
taking the $\Lambda_\tau \to \infty$ limit. To achieve our goal
of deriving nuclear potentials, we first performed integration over the
pion fields in the path integral, yielding a nonlocal
Lagrangian for nucleons.
In the second step, we applied nonlocal redefinitions of the nucleon fields, chosen
in such a way that the interaction part of the transformed Lagrangian
acquires an instantaneous form. Nuclear
forces can then be directly read off from the Lagrangian. It is, however, 
important to keep in mind that nonlocal field redefinitions also induce 
nontrivial contributions to the functional determinant, which need to be
accounted for and, in fact, give rise to loop contributions to
the nuclear potentials in the considered formalism. We have shown how the functional determinant can be
evaluated using the $\zeta$-function regularization.

Our analysis in
this paper is restricted to the contributions of second order in powers of the
coupling constant $g$. We have shown that our method leads to the 
correct one-pion exchange potential and also calculated the order-$g^2$
correction to the nucleon self-energy generated by the functional
determinant, thereby reproducing the well-known leading non-analytic
contribution to the nucleon mass (after performing the
$1/\Lambda$-expansion). On the other hand, the formulation we
employ does, per construction,
not produce corrections to the nucleon $Z$-factor, i.e., one has
$Z=1$ to all orders in $g$. This is not an issue since $Z$-factors are
scheme-dependent unobservable quantities. Nevertheless, for the
sake of completeness and as an additional check of our method, we have
shown how the  original expression for
the $Z$-factor can be restored by explicitly keeping in the path
integral terms induced by nucleon field redefinitions that  involve
external sources. 

The method we have presented is rather  general and particularly well suited 
for the  derivation of regularized nuclear potentials in the famework
of chiral EFT. In this paper we have limited ourselves to a simple model, which does not
feature nontrivial symmetry aspects. In the realistic case of
chiral EFT, regularization of the pion
field will have  to be performed
in such a way that the
chiral and gauge symmetries are preserved. In a subsequent publication
\cite{KE_ToAppear}, we will show how this can be achieved in an elegant way using
the gradient flow method. Our new path integral approach can
then be applied to the regularized chiral Lagrangian in order to derive nuclear forces and currents 
by utilizing the standard power counting of
ChPT.  This opens the way for pushing the accuracy
of chiral EFT for few-nucleon systems and for processes
involving external probes to N$^3$LO and beyond.
Last but not least,  we believe that the approach of reducing
quantum field theory to a quantum mechanical problem
presented here may also be useful for other applications such as,
e.g., potential nonrelativistic EFTs of QCD
and QED \cite{Caswell:1985ui,Pineda:1997bj,Brambilla:2004jw}.

\section*{Acknowledgments}

We are grateful to Vadim Baru, Arseniy Filin, Ashot Gasparyan, Jambul
Gegelia and Ulf~Mei{\ss}ner for useful comments on the manuscript. 
We also thank all members of the LENPIC collaboration for
sharing their insights into the considered topics.
This work is supported in part
by the European Research Council (ERC) under the EU
Horizon 2020 research and innovation programme (ERC AdG
NuclearTheory, grant agreement No. 885150), by DFG and NSFC through
funds provided to the Sino-German CRC 110 ``Symmetries and the
Emergence of Structure in QCD'' (DFG Project ID 196253076 - TRR 110, NSFC Grant No. 11621131001), by the MKW NRW under the funding code NW21-024-A and by
the EU Horizon 2020 research and innovation programme (STRONG-2020,
grant agreement No. 824093).

\appendix
\renewcommand{\theequation}{\thesection.\arabic{equation}}

\section{The functional determinant}
\label{functional:det}
In this Appendix we provide, for the sake of completeness,  a detailed derivation of some (well known)
results for the functional determinant~\cite{ZinnJustin}. We consider a path-integral over Grassmann variables
\beqa
\int D N D N^\dagger F[N, N^\dagger ] \,,
\eeqa
and perform some field transformation 
\beqa
N&\to& N^\prime(N,N^\dagger),\quad N^\dagger\to
N^{\prime\,\dagger}(N,N^\dagger)\,.
\eeqa
The new fields $N^\prime$ and $N^{\prime\,\dagger}$ are assumed to be
given by a sum of odd products of Grassmann fields $N$ and $N^\dagger$,
i.e., 
denoting $N$ and $N^\dagger$ ($N^\prime$ and $N^{\prime\,\dagger}$)
collectively by $\theta_j$ ($\theta_j^\prime$), we have: 
\beqa
\theta_j^\prime&=&\sum_{k=0}^n\sum_{L_k}
c_{j,L_k}\prod_{i=1}^{2k+1}\theta_{l_i},\quad j=1,\dots,n\,,\label{odd:number:Grassmann}
\eeqa
where $L_k=\{l_1,\dots, l_k\}$ and
\beqa
\sum_{L_k}&\equiv&\sum_{l_1=1}^{n}\dots\sum_{l_k=1}^n\,,
\eeqa
with $c_{j,L_k}\in\mathbb{C}$.
Furthermore, we assume that the field transformation is invertible such that
we can also describe $\theta$ as a function of $\theta^\prime$. We
want to show the relationship
\beqa
\int D \theta F[\theta]&=&\int D \theta^\prime \, {\rm Det}\left[
\frac{\delta \theta^\prime}{\delta \theta} 
\right] F[\theta(\theta^\prime)]\,.\label{functional:determinant:statement}
\eeqa
We first make use of the equivalence of the operations of differentiation and
integration for Grassmann variables:
\beqa
\int D \theta F[\theta]&=&\frac{\partial}{\partial
  \theta_1}\dots\frac{\partial}{\partial \theta_n} F[\theta].
\eeqa
Using a chain-rule for a change of variables we obtain
\beqa
\frac{\partial}{\partial
  \theta_1}\dots\frac{\partial}{\partial \theta_n}
F[\theta(\theta^\prime)]
&=&\sum_{j_{1},\dots,j_n=1}^n\frac{\partial\theta_{j_{1}}^\prime}{\partial\theta_{1}}\dots\frac{\partial\theta_{j_n}^\prime}{\partial\theta_n}
\frac{\partial}{\partial\theta_{j_1}^\prime}\dots\frac{\partial}{\partial
\theta_{j_n}^\prime}
F[\theta(\theta^\prime)]\nn
&=&\sum_{j_{1},\dots,j_n=1}^n\frac{\partial\theta_{j_{1}}^\prime}{\partial\theta_{1}}\dots\frac{\partial\theta_{j_n}^\prime}{\partial\theta_n}\epsilon_{j_1,\dots,j_n}
\frac{\partial}{\partial\theta_{1}^\prime}\dots\frac{\partial}{\partial
\theta_{n}^\prime}
F[\theta(\theta^\prime)]\,,
\eeqa
where in the last step we used the fact that $(\partial/\partial
\theta_j)^n=0$ for $n>1$,
and for this reason only derivatives with different indices $j_1\neq
j_2\neq\dots\neq j_n$ will contribute to a non-vanishing result. Due to
Grassmann-nature of $\theta^\prime$-variables, all derivatives in
$\theta^\prime$ anticomute and we obtain
\beqa
\frac{\partial}{\partial\theta_{j_1}^\prime}\dots\frac{\partial}{\partial
\theta_{j_n}^\prime}
F[\theta(\theta^\prime)]&=&\epsilon_{j_1,\dots,j_n}
\frac{\partial}{\partial\theta_{1}^\prime}\dots\frac{\partial}{\partial
\theta_{n}^\prime}
F[\theta(\theta^\prime)]\,.
\eeqa
The relationship in 
Eq.~(\ref{functional:determinant:statement}) is then obtained by using
the definition of a determinant 
\beqa
 {\rm Det}\left[
\frac{\delta \theta^\prime}{\delta \theta} 
\right]=\sum_{j_1,\dots,j_n=1}^{n}\epsilon_{j_1,\dots,j_n}\frac{\partial\theta_{j_{1}}^\prime}{\partial\theta_{1}}\dots\frac{\partial\theta_{j_n}^\prime}{\partial\theta_n}\,.
\eeqa

\section{$\zeta$-function regularization of the functional
  determinant}
\label{zetafunctionReg}
In this appendix we calculate the functional determinant within the
$\zeta$-function regularization approach. Let $A$ be an operator with
a complete discrete set of quadratintegrable eigenfunctions
$\{f_n(x)\}$ with real positive eigenvalues $\{\lambda_n\}$ such that its
spectrum is bounded from below. The mass-dimension of the operator $A$
is assumed to be zero. For the case of continuous spectrum we
put the system in a finite box, which makes the spectrum discrete. At
the end of the calculation, we will take the infinite-volume limit. The Riemann
$\zeta$-function is defined by
\beqa
\zeta:\mathbb{C}&\to&\mathbb{C}\nn
s&\mapsto&\bigg\{\begin{array}{l}
\sum_{n=1}^\infty\frac{1}{n^s}\quad {\rm for\, Re\,}s>1\,,\\[4pt]
{\rm analytic\,continuation\;from\;the \;Re\,}s>1\;{\rm area} \,.
\end{array}
\eeqa
We define the generalized $\zeta_A$-function from the spectrum of the
operator $A$ via~\cite{ramond2007field,Meissner:2022cbi}
\beqa
\zeta_A(s)&=&
\sum_{n=1}^\infty\frac{1}{\lambda_n^s}\,.\label{zetaADefinition}
\eeqa
We assume that the sum in Eq.~(\ref{zetaADefinition}) converges for
$s$ in
some area of the complex plane and perform an analytic continuation
for other values of $s$. Taking derivative in $s$ and assuming that
the operations of 
summation and differentiation can be interchanged we obtain
\beqa
\frac{d\zeta_A}{d s}&=&\sum_{n=1}^{\infty}\frac{d}{d s}e^{-s\log
  \lambda_n}\,=\,-\sum_{n=1}^\infty\frac{\log \lambda_n}{\lambda_n^s}.
\eeqa
After an analytic continuation to $s=0^+$, we have
\beqa
\label{AppBtemp1}
\lim_{s\to 0^+}\frac{d\zeta_A}{d s}&=&-\sum_{n=1}^\infty
\log\,\lambda_n\,=\,-{\rm Tr}\,\log\,A,
\eeqa
where $0^+$ means that when performing the
analytic continuation we have constrained the values of $s$ to  ${\rm
  Re}\, s>0$.
Thus, a task of evaluating the determinant of an operator $A$
reduces to the calculation of the corresponding generalized $\zeta_A$-function.

To compute the $\zeta_A$-function we
define the generalized heat kernel via
\beqa
G_A(x,a;y,b;\tau)&=&\sum_{n=1}^\infty
e^{-\lambda_n\,\tau}f_n(x,a)f_n^*(y,b)\,=\,\langle x,a|e^{-\tau\,A}|y,b\rangle,
\eeqa
where $\tau>0$, $x$, $y$ denote the coordinates while $a$, $b$ are
discrete quantum numbers like spin or isospin. Further, $f_n(x,a)$ refer to the
coordinate-space representation of an eigenstate $|n>$  of the operator $A$:
\beq
f_n(x,a) \; =\;  \langle x,a|n\rangle\,, \quad \quad 
A|n\rangle \; =\; \lambda_n|n\rangle\,.
\eeq
The heat kernel $G_A$ satisfies the initial condition
\beqa \lim_{\tau\to 0^+} G_A(x,a;y,b;\tau)&=&\sum_{n=1}^\infty
f_n(x,a) f_n^*(y,b)\,=\,\delta(x-y)\, \delta_{ab},
\eeqa
where in the last step we used the completeness relation of
eigenfunctions $\{f_n\}$. In four-dimensional Euclidean space-time
we obtain
\beqa
\label{A2temp1}
\sum_a\int d^4 x\lim_{y\to x} G_A(x, a;y,a;\tau)&=&\sum_{n=1}^\infty
e^{-\lambda_n\,\tau} \sum_a\int d^4 x f_n(x,a) f_n^*(x,a)\,=\, \sum_{n=1}^\infty
e^{-\lambda_n\,\tau} \,=\,{\rm Tr}\,e^{-\tau\,A},
\eeqa
where we used orthonormalization condition
\beqa
\sum_a\int d^4 x\, f_n^*(x,a) f_m(x,a)&=&\delta_{n m}.
\eeqa
For ${\rm Re}\,s>0$ and $\lambda_n>0$, we can use the relationship
\beqa
\int_0^\infty d\tau \,\tau^{s-1} e^{-\lambda_n\,\tau}&=&\frac{\Gamma(s)}{\lambda_n^s},\label{GammaFWithan}
\eeqa
where the Euler Gamma-function is defined by
\beqa
\Gamma(s)&=&\int_0^\infty d\tau\,\tau^{s-1} e^{-\tau}.
\eeqa
With Eq.~(\ref{GammaFWithan}) we obtain from Eq.~(\ref{A2temp1}):
\beqa
\int_0^\infty d\tau\,\tau^{s-1}\sum_a\int d^4 x\lim_{y\to x}
G_A(x,a;y,a;\tau)\,=\,\sum_{n=1}^\infty \frac{\Gamma(s)}{\lambda_n^s}.
\eeqa
Thus, the $\zeta_A$-function can be expressed in
terms of the generalized heat kernel via
\beqa
\zeta_A(s)&=&\frac{1}{\Gamma(s)}\int_0^\infty d\tau\,\tau^{s-1}\sum_a\int
d^4 x \,\lim_{y\to x}G_A(x,a;y,a;\tau).\label{zetaViaHeatKernel}
\eeqa
This suggests the following procedure. We first transform everything
to momentum space to obtain
\beqa
\tilde G_A(p,a;q,b;\tau)&=&\sum_{c,d}\int d^4 x\,d^4 y\langle p,a| x,c\rangle\langle x,c|
e^{-A\,\tau}|y,d\rangle\langle y,d|q,b\rangle\,=\,\langle p,a|e^{-A\,\tau}| q,b\rangle\,,
\eeqa
and then calculate the operator $e^{-\tau A}$ using time-dependent
perturbation theory. To this aim, we decompose the operator $A=A_0+\delta A$ into its
unperturbed part, $A_0=\big(-\partial^2+M^2\big)/\mu^2$, and the perturbation $\delta
A$. The renormalization scale $\mu$ is introduced here to make $A_0$
dimensionless.  Note that for the case at hand, $\delta A$ is a functional of
Grassman-variables (nucleon fields). We, therefore, can not claim positive
definiteness of  $A$, but the operator $A_0$ has indeed only positive spectrum. We 
formaly extend Eq.~(\ref{AppBtemp1}) to Grassmann-valued functionals
and perform an expansion in $\delta A$. To do this we use the well-known relationship for the proper-time-evolution
operator in the interaction picture~\cite{Feynman:1951gn}
\beqa
e^{\tau A_0} e^{-\tau A}&=&T\exp\bigg(-\int_0^\tau ds\,\delta A_I(s)\bigg),\label{ProperTimeEvolutionIntPicture}
\eeqa
where $T$ is the proper-time-ordering operator and 
\beqa
\delta A_I(\tau)&=& e^{\tau A_0}\delta A e^{-\tau A_0}.
\eeqa
The proof of this result is very simple. The left-hand side $O(\tau)=e^{\tau
  A_0} e^{-\tau A}$ of
Eq.~(\ref{ProperTimeEvolutionIntPicture}) satisfies a first-order
differential equation in the proper time,
\beqa
\frac{\partial}{\partial \tau} O(\tau)&=&-\delta A_I(\tau) O(\tau)\,,\label{DifferentialEqOfTexp}
\eeqa
with the boundary condition 
\beqa
O(0)&=&1.\label{DifferentialEqOfTexpBoundary}
\eeqa
The rhs of Eq.~(\ref{ProperTimeEvolutionIntPicture}) satisfies the same
differential equation Eq.~(\ref{DifferentialEqOfTexp}) with the same
boundary condition in Eq.~(\ref{DifferentialEqOfTexpBoundary}).
Eq.~(\ref{ProperTimeEvolutionIntPicture}) then immediately follows from the uniqueness of
the solution to Eqs.~(\ref{DifferentialEqOfTexp})), (\ref{DifferentialEqOfTexpBoundary}).  

We calculate the right-hand side of
Eq.~(\ref{ProperTimeEvolutionIntPicture}) by expanding the exponential
function in $\delta A_I$:
\beqa
e^{-\tau A}=e^{-\tau A_0} \sum_{n=0}^\infty (-1)^n\int_0^\tau
ds_1\dots
\int_0^{s_{n-1}} ds_n \delta A_I(s_1)\delta A_I(s_2)\cdots\delta A_I(s_n).
\eeqa
Up to third order in $\delta A_I$, we find
\beqa
\tilde G_A(p,a;q,b;\tau)&=&e^{-\tau\big(p^2+M^2\big)/\mu^2}\bigg[(2\pi)^4\delta(p-q)\delta_{ab}\;-\;\int_0^\tau
ds\, e^{s\big(p^2-q^2\big)/\mu^2}\langle p,a|\delta A|q,b\rangle \nn
&+&\sum_c\int \frac{d^4
  k}{(2\pi)^4}
\int_0^\tau ds_1\int_0^{s_1} ds_2
\,e^{s_1\big(p^2-k^2\big)/\mu^2+s_2\big(k^2-q^2\big)/\mu^2}
\langle p,a|\delta
A|k,c\rangle\langle k,c|\delta A|q,b\rangle \nn
&-& \sum_{c_1,c_2}\int \frac{d^4
  k_1}{(2\pi)^4} \frac{d^4 k_2}{(2\pi)^4}
\int_0^\tau ds_1\int_0^{s_1} ds_2\int_0^{s_2} ds_3
\,e^{s_1\big(p^2-k_1^2\big)/\mu^2+s_2\big(k_1^2-k_2^2\big)/\mu^2+s_3\big(k_2^2-q^2\big)/\mu^2}\nn
&\times&\langle p,a|\delta
A|k_1,c_1\rangle\langle k_1,c_1|\delta A|k_2,c_2\rangle\langle
k_2,c_2|\delta A|q,b\rangle \;+\;{\cal O}\big(\delta A_I^4\big)\bigg].
\eeqa
Notice that there 
is no exponential increase in the integrals over intermediate
momenta, such that all integrals in the above equation are convergent
(provided the operator $\delta A$ is  well behaved). 
For example, the second-order expression has the form
\beqa
\int_0^\tau ds_1\int_0^{s_1} ds_2\int\frac{d^4 k}{(2\pi)^4}\langle p,a|\delta
A_I(s_1)|k,c\rangle\langle k,c|\delta A_I(s_2)|q,b\rangle
&=&\sum_c\int \frac{d^4
  k}{(2\pi)^4}
\int_0^\tau ds_1\int_0^{s_1} ds_2
\,e^{s_1\big(p^2-k^2\big)/\mu^2+s_2\big(k^2-q^2\big)/\mu^2}\nn
&\times &
\langle p,a|\delta
A|k,c\rangle\langle k,c|\delta A|q,b\rangle.\label{secondOrderSymbolicaly}
\eeqa 
Since the integration variables in Eq.(\ref{secondOrderSymbolicaly})
fulfill $s_1\geq s_2$, there are no exponentially growing factors in
the momentum $k$.

To obtain the $\zeta_A$-function given in Eq.~(\ref{zetaViaHeatKernel}),
we perform the integration over the coordinates:
\beqa
\sum_a\int d^4x \lim_{y\to x} G_A(x,a,y,a;\tau)&=&\sum_a\int d^4 x \int \frac{d^4
  p}{(2\pi)^4} \frac{d^4 q}{(2\pi)^4}e^{i(q-p)\cdot x} \tilde
G_(p,a;q,a;\tau)\nn
&=&
\int \frac{d^4p}{(2\pi)^4} \tilde G_A(p,a;p,a;\tau)\nn
&=&\sum_a\int \frac{d^4 p}{(2\pi)^4}e^{-\tau
  \big(p^2+M^2\big)/\mu^2}\bigg[V \; - \;\tau\langle p,a|\delta A|
p,a\rangle\nn
&
+&\sum_c\int \frac{d^4
  k}{(2\pi)^4}\int_0^\tau ds_1\int_0^{s_1} ds_2
\,e^{s_1\big(p^2-k^2\big)/\mu^2+s_2\big(k^2-p^2\big)/\mu^2}
\langle p,a|\delta
A|k,c\rangle\langle k,c|\delta A|p,a\rangle \nn
&-&\sum_{c_1,c_2}\int \frac{d^4 k_1}{(2\pi)^4}\frac{d^4 k_2}{(2\pi)^4}\int_0^\tau ds_1\int_0^{s_1} ds_2\int_0^{s_2} ds_3
\,e^{s_1\big(p^2-k_1^2\big)/\mu^2+s_2\big(k_1^2-k_2^2\big)+s_3\big(k_2^2-p^2\big)/\mu^2}\nn
&&\times\langle p,a|\delta
A|k_1,c_1\rangle\langle k_1,c_1|\delta A|k_2,c_2\rangle\langle k_2,c_2|\delta A|p,a\rangle\;+\;{\cal
  O}\big(\delta A^4\big)\bigg],
\eeqa
where $V=\int d^4 x$ is the space-time volume. Inserting this result
into Eq.~(\ref{zetaViaHeatKernel}) we finally obtain
\beqa
\zeta_A(s)&=&\int_0^\infty d\tau\frac{\tau^{s-1}}{\Gamma(s)}\sum_a\int \frac{d^4
  p}{(2\pi)^4}e^{-\tau\big(p^2+M^2\big)/\mu^2}\bigg[V\;-\;\tau\langle
p,a|\delta A|p,a\rangle \nn
&+&\sum_c\int \frac{d^4
  k}{(2\pi)^4}\int_0^\tau ds_1\int_0^{s_1} ds_2 e^{s_1\big(p^2-k^2\big)/\mu^2+s_2\big(k^2-p^2\big)/\mu^2}\langle p,a|\delta
A|k,c\rangle\langle k,c|\delta A|p,a\rangle\nn
&-&\sum_{c_1,c_2}\int \frac{d^4
  k_1}{(2\pi)^4}\frac{d^4 k_2}{(2\pi)^4} \int_0^\tau ds_1\int_0^{s_1}
ds_2\int_0^{s_2} ds_3
e^{s_1\big(p^2-k_1^2\big)/\mu^2+s_2\big(k_1^2-k_2^2\big)/\mu^2+s_3\big(k_2^2-p^2\big)/\mu^2}\nn
&\times&\langle p,a|\delta
A|k_1,c_1\rangle\langle k_1,c_1|\delta A|k_2,c_2\rangle  \langle
k_2,c_2|\delta A|p,a\rangle \;+ \;{\cal
  O}\big(\delta A^4\big)\bigg].\label{ZetaAFunctionMomentumSpace}
\eeqa
Notice that the complex parameter $s$ plays the role of a regulator.

As a
consistency check, we can assume that the operator $\delta A$ is well
behaved, so that one can calculate  the derivative of $\zeta_A(s)$
and take the limit $s \to 0^+$ prior to performing integrations
over momenta.
In this case we expect to obtain the standard
result
\beqa
\zeta^\prime_A(0)-\zeta^\prime_{A_0}(0)&=&-{\rm
  Tr}\log\bigg(1+\frac{\delta
  A}{A_0}\bigg)\;=\;-\sum_a\int\frac{d^4p}{(2\pi)^4}\frac{\mu^2}{p^2+M^2}\langle
p, a|\delta A| p , a\rangle \nn
&+&\frac{1}{2}\sum_{a,c}\int \frac{d^4
  p}{(2\pi)^4}\frac{d^4 k}{(2\pi)^4}\frac{\mu^2}{p^2+M^2}\frac{\mu^2}{k^2+M^2}
\langle
p, a|\delta A| k , c\rangle \langle k, c|\delta A| p , a\rangle \;+\; {\cal
  O}\big(\delta A^3\big).\label{ZetaDifferenceNaiv}
\eeqa
To see if we can reproduce this result, we perform the integrations over
$s_1, s_2$ and $\tau$ in Eq.~\ref{ZetaAFunctionMomentumSpace}. After
taking the derivative in $s$ and setting $s = 0^+$, we obtain
\beqa
\zeta_A^\prime(0)-\zeta_{A_0}^\prime(0)&=&\frac{\partial}{\partial s}\bigg|_{s=0^+}\sum_a\int \frac{d^4
  p}{(2\pi)^4}\frac{\mu^{2 s}}{\big(p^2+M^2\big)^s}\bigg\{-\frac{s \,\mu^2}{p^2+M^2}\langle
p,a|\delta A|p,a\rangle \nn
&+&\sum_c\int \frac{d^4
  k}{(2\pi)^4}\frac{\mu^4}{(p^2-k^2)^2}\bigg[\bigg(\frac{p^2+M^2}{k^2+M^2}\bigg)^s-1-s\frac{p^2-k^2}{p^2+M^2}\bigg]\langle p,a|\delta
A|k,c\rangle\langle k,c|\delta A|p,a\rangle
\;+\;{\cal
  O}\big(\delta A^3\big)\bigg\}\nn
&=&\sum_a\int \frac{d^4
  p}{(2\pi)^4}\bigg\{-\frac{\mu^2}{p^2+M^2}\langle
p,a|\delta A|p,a\rangle 
+\sum_c\int \frac{d^4
  k}{(2\pi)^4}\frac{\mu^4}{(p^2-k^2)^2}\bigg[\log\bigg(\frac{p^2+M^2}{k^2+M^2}\bigg)-\frac{p^2-k^2}{p^2+M^2}\bigg]\nn
&\times&\langle p,a|\delta
A|k,c\rangle\langle k,c|\delta A|p,a\rangle
\;+\;{\cal
  O}\big(\delta A^3\big)\bigg\}.\label{ZetaAFunctionMomentumSpacePrepare}
\eeqa
The second-order term in Eq.~(\ref{ZetaAFunctionMomentumSpacePrepare}) is not
identical to the corresponding term in Eq.~(\ref{ZetaDifferenceNaiv}). However,
a symmetrization of the integrand with respect to the variables
$(p,a,k,c)\leftrightarrow (k,c,p,a)$
does not change the value of the integral. Thus, we can replace the last term in
Eq.~(\ref{ZetaAFunctionMomentumSpacePrepare}) by
\beqa
\frac{\mu^4}{(p^2-k^2)^2}\bigg[\log\bigg(\frac{p^2+M^2}{k^2+M^2}\bigg)-\frac{p^2-k^2}{p^2+M^2}\bigg]&\to&\frac{1}{2}\frac{\mu^4}{(p^2-k^2)^2}\bigg[\frac{p^2-k^2}{k^2+M^2}-\frac{p^2-k^2}{p^2+M^2}\bigg]\;=\; \frac{1}{2}\frac{\mu^2}{p^2+M^2}\frac{\mu^2}{k^2+M^2},\nonumber
\eeqa
so that Eq.~(\ref{ZetaDifferenceNaiv}) is indeed reproduced.
Notice that for the case at hand, the operator
$\delta A$ will not provide enough suppression at high momenta, so
that we continue with Eq.~(\ref{ZetaAFunctionMomentumSpace}) and take
the derivative at $s=0$ after integrating over momenta and over the
parameters $\tau$, $s_1$ and $s_2$.


\begin{thebibliography}{99}


\bibitem{Weinberg:1990rz}
S.~Weinberg,
Phys. Lett. B \textbf{251}, 288-292 (1990).

\bibitem{Weinberg:1991um}
S.~Weinberg,
Nucl. Phys. B \textbf{363}, 3-18 (1991).

\bibitem{Epelbaum:2008ga}
E.~Epelbaum, H.~W.-Hammer and U.-G.~Mei{\ss}ner,
Rev. Mod. Phys. \textbf{81}, 1773-1825 (2009)
[arXiv:0811.1338 [nucl-th]].

\bibitem{Machleidt:2011zz}
R.~Machleidt and D.~R.~Entem,
Phys. Rept. \textbf{503}, 1-75 (2011)
[arXiv:1105.2919 [nucl-th]].

\bibitem{Bernard:1995dp}
V.~Bernard, N.~Kaiser and U.-G.~Mei{\ss}ner,
Int. J. Mod. Phys. E \textbf{4}, 193-346 (1995)
[arXiv:hep-ph/9501384 [hep-ph]].


\bibitem{Bernard:2007zu}
V.~Bernard,
Prog. Part. Nucl. Phys. \textbf{60}, 82-160 (2008)
[arXiv:0706.0312 [hep-ph]].

\bibitem{Scherer:2012xha}
S.~Scherer and M.~R.~Schindler,
Lect. Notes Phys. \textbf{830}, pp.1-338 (2012)

\bibitem{Ordonez:1995rz}
C.~Ordonez, L.~Ray and U.~van Kolck,
Phys. Rev. C \textbf{53}, 2086-2105 (1996)
[arXiv:hep-ph/9511380 [hep-ph]].


\bibitem{Pastore:2008ui}
S.~Pastore, R.~Schiavilla and J.~L.~Goity,
Phys. Rev. C \textbf{78}, 064002 (2008)
[arXiv:0810.1941 [nucl-th]].

\bibitem{Pastore:2009is}
S.~Pastore, L.~Girlanda, R.~Schiavilla, M.~Viviani and R.~B.~Wiringa,
Phys. Rev. C \textbf{80}, 034004 (2009)
[arXiv:0906.1800 [nucl-th]].

\bibitem{Pastore:2011ip}
S.~Pastore, L.~Girlanda, R.~Schiavilla and M.~Viviani,
Phys. Rev. C \textbf{84}, 024001 (2011)
[arXiv:1106.4539 [nucl-th]].

\bibitem{Baroni:2015uza}
A.~Baroni, L.~Girlanda, S.~Pastore, R.~Schiavilla and M.~Viviani,
Phys. Rev. C \textbf{93}, no.1, 015501 (2016)
[erratum: Phys. Rev. C \textbf{93}, no.4, 049902 (2016); erratum: Phys. Rev. C \textbf{95}, no.5, 059901 (2017)]
[arXiv:1509.07039 [nucl-th]].

\bibitem{deVries:2020iea}
J.~de Vries, E.~Epelbaum, L.~Girlanda, A.~Gnech, E.~Mereghetti and M.~Viviani,
Front. in Phys. \textbf{8}, 218 (2020)
[arXiv:2001.09050 [nucl-th]].

\bibitem{Baru:2019ndr}
V.~Baru, E.~Epelbaum, J.~Gegelia and X.~L.~Ren,
Phys. Lett. B \textbf{798}, 134987 (2019)
[arXiv:1905.02116 [nucl-th]].


\bibitem{Epelbaum:1998ka}
E.~Epelbaum, W.~Gl\"ockle and U.-G.~Mei{\ss}ner,
Nucl. Phys. A \textbf{637}, 107-134 (1998)
[arXiv:nucl-th/9801064 [nucl-th]].

\bibitem{Epelbaum:1999dj}
E.~Epelbaum, W.~Gl\"ockle and U.~G.~Mei{\ss}ner,
Nucl. Phys. A \textbf{671}, 295-331 (2000)
[arXiv:nucl-th/9910064 [nucl-th]].

\bibitem{Epelbaum:2002gb}
E.~Epelbaum, U.-G.~Mei{\ss}ner and W.~Gl\"ockle,
Nucl. Phys. A \textbf{714}, 535-574 (2003)
[arXiv:nucl-th/0207089 [nucl-th]].

\bibitem{Epelbaum:2005fd}
E.~Epelbaum and U.-G.~Mei{\ss}ner,
Phys. Rev. C \textbf{72}, 044001 (2005)
[arXiv:nucl-th/0502052 [nucl-th]].

\bibitem{Epelbaum:2005bjv}
E.~Epelbaum,
Phys. Lett. B \textbf{639}, 456-461 (2006)
[arXiv:nucl-th/0511025 [nucl-th]].

\bibitem{Epelbaum:2007us}
E.~Epelbaum,
Eur. Phys. J. A \textbf{34}, 197-214 (2007)
[arXiv:0710.4250 [nucl-th]].

\bibitem{Bernard:2007sp}
V.~Bernard, E.~Epelbaum, H.~Krebs and U.-G.~Mei{\ss}ner,
Phys. Rev. C \textbf{77}, 064004 (2008)
[arXiv:0712.1967 [nucl-th]].

\bibitem{Bernard:2011zr}
V.~Bernard, E.~Epelbaum, H.~Krebs and U.-G.~Mei{\ss}ner,
Phys. Rev. C \textbf{84}, 054001 (2011)
[arXiv:1108.3816 [nucl-th]].

\bibitem{Krebs:2012yv}
H.~Krebs, A.~Gasparyan and E.~Epelbaum,
Phys. Rev. C \textbf{85}, 054006 (2012)
[arXiv:1203.0067 [nucl-th]].

\bibitem{Krebs:2013kha}
H.~Krebs, A.~Gasparyan and E.~Epelbaum,
Phys. Rev. C \textbf{87}, no.5, 054007 (2013)
[arXiv:1302.2872 [nucl-th]].

\bibitem{Kolling:2009iq}
S.~K\"olling, E.~Epelbaum, H.~Krebs and U.-G.~Mei{\ss}ner,
Phys. Rev. C \textbf{80}, 045502 (2009)
[arXiv:0907.3437 [nucl-th]].

\bibitem{Kolling:2011mt}
S.~K\"olling, E.~Epelbaum, H.~Krebs and U.-G.~Mei{\ss}ner,
Phys. Rev. C \textbf{84}, 054008 (2011)
[arXiv:1107.0602 [nucl-th]].

\bibitem{Krebs:2016rqz}
H.~Krebs, E.~Epelbaum and U.-G.~Mei\ss{}ner,
Annals Phys. \textbf{378}, 317-395 (2017)
[arXiv:1610.03569 [nucl-th]].

\bibitem{Krebs:2019aka}
H.~Krebs, E.~Epelbaum and U.-G.~Mei\ss{}ner,
Few Body Syst. \textbf{60}, no.2, 31 (2019)
[arXiv:1902.06839 [nucl-th]].

\bibitem{Krebs:2020plh}
H.~Krebs, E.~Epelbaum and U.-G.~Mei\ss{}ner,
Eur. Phys. J. A \textbf{56}, no.9, 240 (2020)
[arXiv:2005.07433 [nucl-th]].

\bibitem{Krebs:2020pii}
H.~Krebs,
Eur. Phys. J. A \textbf{56}, no.9, 234 (2020)
[arXiv:2008.00974 [nucl-th]].

\bibitem{Kaiser:1997mw}
N.~Kaiser, R.~Brockmann and W.~Weise,
Nucl. Phys. A \textbf{625}, 758-788 (1997)
[arXiv:nucl-th/9706045 [nucl-th]].

\bibitem{Kaiser:1998wa}
N.~Kaiser, S.~Gerstendorfer and W.~Weise,
Nucl. Phys. A \textbf{637}, 395-420 (1998)
[arXiv:nucl-th/9802071 [nucl-th]].

\bibitem{Kaiser:1999ff}
N.~Kaiser,
Phys. Rev. C \textbf{61}, 014003 (2000)
[arXiv:nucl-th/9910044 [nucl-th]].

\bibitem{Kaiser:1999jg}
N.~Kaiser,
Phys. Rev. C \textbf{62}, 024001 (2000)
[arXiv:nucl-th/9912054 [nucl-th]].

\bibitem{Kaiser:2001dm}
N.~Kaiser,
Phys. Rev. C \textbf{63}, 044010 (2001)
[arXiv:nucl-th/0101052 [nucl-th]].

\bibitem{Kaiser:2001pc}
N.~Kaiser,
Phys. Rev. C \textbf{64}, 057001 (2001)
[arXiv:nucl-th/0107064 [nucl-th]].

\bibitem{Kaiser:2001at}
N.~Kaiser,
Phys. Rev. C \textbf{65}, 017001 (2002)
[arXiv:nucl-th/0109071 [nucl-th]].

\bibitem{Entem:2015xwa}
D.~R.~Entem, N.~Kaiser, R.~Machleidt and Y.~Nosyk,
Phys. Rev. C \textbf{92}, no.6, 064001 (2015)
[arXiv:1505.03562 [nucl-th]].

\bibitem{Epelbaum:2002vt}
E.~Epelbaum, A.~Nogga, W.~Gloeckle, H.~Kamada, U.-G.~Mei{\ss}ner and H.~Witala,
Phys. Rev. C \textbf{66}, 064001 (2002)
[arXiv:nucl-th/0208023 [nucl-th]].

\bibitem{Epelbaum:2003gr}
E.~Epelbaum, W.~Gl\"ockle and U.-G.~Mei{\ss}ner,
Eur. Phys. J. A \textbf{19}, 125-137 (2004)
[arXiv:nucl-th/0304037 [nucl-th]].

\bibitem{Epelbaum:2003xx}
E.~Epelbaum, W.~Gl\"ockle and U.-G.~Mei{\ss}ner,
Eur. Phys. J. A \textbf{19}, 401-412 (2004)
[arXiv:nucl-th/0308010 [nucl-th]].

\bibitem{Epelbaum:2004fk}
E.~Epelbaum, W.~Gl\"ockle and U.-G.~Mei{\ss}ner,
Nucl. Phys. A \textbf{747}, 362-424 (2005)
[arXiv:nucl-th/0405048 [nucl-th]].

\bibitem{Entem:2003ft}
D.~R.~Entem and R.~Machleidt,
Phys. Rev. C \textbf{68}, 041001 (2003)
[arXiv:nucl-th/0304018 [nucl-th]].


\bibitem{Gezerlis:2013ipa}
A.~Gezerlis, I.~Tews, E.~Epelbaum, S.~Gandolfi, K.~Hebeler, A.~Nogga and A.~Schwenk,
Phys. Rev. Lett. \textbf{111}, no.3, 032501 (2013)
[arXiv:1303.6243 [nucl-th]].

\bibitem{Ekstrom:2013kea}
A.~Ekstr\"om, G.~Baardsen, C.~Forss\'en, G.~Hagen, M.~Hjorth-Jensen, G.~R.~Jansen, R.~Machleidt, W.~Nazarewicz, T.~Papenbrock and J.~Sarich, \textit{et al.}
Phys. Rev. Lett. \textbf{110}, no.19, 192502 (2013)
[arXiv:1303.4674 [nucl-th]].

\bibitem{Piarulli:2014bda}
M.~Piarulli, L.~Girlanda, R.~Schiavilla, R.~Navarro P\'erez, J.~E.~Amaro and E.~Ruiz Arriola,
Phys. Rev. C \textbf{91}, no.2, 024003 (2015)
[arXiv:1412.6446 [nucl-th]].

\bibitem{Epelbaum:2014efa}
E.~Epelbaum, H.~Krebs and U.-G.~Mei\ss{}ner,
Eur. Phys. J. A \textbf{51}, no.5, 53 (2015)
[arXiv:1412.0142 [nucl-th]].

\bibitem{Epelbaum:2014sza}
E.~Epelbaum, H.~Krebs and U.-G.~Mei\ss{}ner,
Phys. Rev. Lett. \textbf{115}, no.12, 122301 (2015)
[arXiv:1412.4623 [nucl-th]].

\bibitem{Ekstrom:2015rta}
A.~Ekstr\"om, G.~R.~Jansen, K.~A.~Wendt, G.~Hagen, T.~Papenbrock, B.~D.~Carlsson, C.~Forss\'en, M.~Hjorth-Jensen, P.~Navr\'atil and W.~Nazarewicz,
Phys. Rev. C \textbf{91}, no.5, 051301 (2015)
[arXiv:1502.04682 [nucl-th]].

\bibitem{Entem:2017gor}
D.~R.~Entem, R.~Machleidt and Y.~Nosyk,
Phys. Rev. C \textbf{96}, no.2, 024004 (2017)
[arXiv:1703.05454 [nucl-th]].

\bibitem{Reinert:2017usi}
P.~Reinert, H.~Krebs and E.~Epelbaum,
Eur. Phys. J. A \textbf{54}, no.5, 86 (2018)
[arXiv:1711.08821 [nucl-th]].

\bibitem{LENPIC:2018ewt}
E.~Epelbaum \textit{et al.} [LENPIC],
Phys. Rev. C \textbf{99}, no.2, 024313 (2019)
[arXiv:1807.02848 [nucl-th]].

\bibitem{Huther:2019ont}
T.~H\"uther, K.~Vobig, K.~Hebeler, R.~Machleidt and R.~Roth,
Phys. Lett. B \textbf{808}, 135651 (2020)
[arXiv:1911.04955 [nucl-th]].

\bibitem{Reinert:2020mcu}
P.~Reinert, H.~Krebs and E.~Epelbaum,
Phys. Rev. Lett. \textbf{126}, no.9, 092501 (2021)
[arXiv:2006.15360 [nucl-th]].

\bibitem{Maris:2020qne}
P.~Maris, E.~Epelbaum, R.~J.~Furnstahl, J.~Golak, K.~Hebeler, T.~H\"uther, H.~Kamada, H.~Krebs, U.-G.~Mei\ss{}ner and J.~A.~Melendez, \textit{et al.}
Phys. Rev. C \textbf{103}, no.5, 054001 (2021)
[arXiv:2012.12396 [nucl-th]].

\bibitem{Somasundaram:2023sup}
R.~Somasundaram, J.~E.~Lynn, L.~Huth, A.~Schwenk and I.~Tews,
[arXiv:2306.13579 [nucl-th]].

\bibitem{Epelbaum:2019kcf}
E.~Epelbaum, H.~Krebs and P.~Reinert,
Front. in Phys. \textbf{8}, 98 (2020)
[arXiv:1911.11875 [nucl-th]].

\bibitem{Krebs:2019uvm}
H.~Krebs,
PoS \textbf{CD2018}, 098 (2019)
[arXiv:1908.01538 [nucl-th]].




\bibitem{Epelbaum:2022cyo}
E.~Epelbaum, H.~Krebs and P.~Reinert,
[arXiv:2206.07072 [nucl-th]].

\bibitem{Slavnov:1971aw}
A.~A.~Slavnov,
Nucl. Phys. B \textbf{31}, 301-315 (1971)

\bibitem{Djukanovic:2004px}
D.~Djukanovic, M.~R.~Schindler, J.~Gegelia and S.~Scherer,
Phys. Rev. D \textbf{72}, 045002 (2005)
[arXiv:hep-ph/0407170 [hep-ph]].

\bibitem{Long:2016vnq}
B.~Long and Y.~Mei,
Phys. Rev. C \textbf{93}, no.4, 044003 (2016)
[arXiv:1605.02153 [nucl-th]].

\bibitem{KE_ToAppear}
H.~Krebs, E.~Epelbaum, {\it Towards consistent nuclear interactions from chiral
  Lagrangians II: Symmetry preserving regularization}, to appear.

\bibitem{Gasser:1983yg}
J.~Gasser and H.~Leutwyler,
Annals Phys. \textbf{158}, 142 (1984)

\bibitem{Jenkins:1990jv}
E.~E.~Jenkins and A.~V.~Manohar,
Phys. Lett. B \textbf{255}, 558-562 (1991)

\bibitem{Bernard:1992qa}
V.~Bernard, N.~Kaiser, J.~Kambor and U.-G.~Mei{\ss}ner,
Nucl. Phys. B \textbf{388}, 315-345 (1992)

\bibitem{Gasparyan:2021edy}
A.~M.~Gasparyan and E.~Epelbaum,
Phys. Rev. C \textbf{105}, no.2, 024001 (2022).

\bibitem{Friar:2004ca}
J.~L.~Friar, U.~van Kolck, M.~C.~M.~Rentmeester and R.~G.~E.~Timmermans,
Phys. Rev. C \textbf{70}, 044001 (2004)
[arXiv:nucl-th/0406026 [nucl-th]].

\bibitem{Epelbaum:2007sq}
E.~Epelbaum, H.~Krebs and U.-G.~Mei{\ss}ner,
Nucl. Phys. A \textbf{806}, 65-78 (2008)
[arXiv:0712.1969 [nucl-th]].

\bibitem{Borasoy:2006qn}
B.~Borasoy, E.~Epelbaum, H.~Krebs, D.~Lee and U.-G.~Mei{\ss}ner,
Eur. Phys. J. A \textbf{31}, 105-123 (2007)
[arXiv:nucl-th/0611087 [nucl-th]].

\bibitem{Haag:1958vt}
R.~Haag,
Phys. Rev. \textbf{112}, 669-673 (1958).

\bibitem{Coleman:1969sm}
S.~R.~Coleman, J.~Wess and B.~Zumino,
Phys. Rev. \textbf{177}, 2239-2247 (1969).


\bibitem{Ecker:1994pi}
G.~Ecker,
Phys. Lett. B \textbf{336}, 508-517 (1994)
[arXiv:hep-ph/9402337 [hep-ph]].

\bibitem{Meissner:1998rw}
U.-G.~Mei{\ss}ner, G.~M\"uller and S.~Steininger,
Annals Phys. \textbf{279}, 1-64 (2000)
[arXiv:hep-ph/9809446 [hep-ph]].

\bibitem{Reuter:1984kw}
M.~Reuter,
Phys. Rev. D \textbf{31}, 1374 (1985)

\bibitem{Gasser:1979hf}
J.~Gasser and A.~Zepeda,
Nucl. Phys. B \textbf{174}, 445 (1980)

\bibitem{Caswell:1985ui}
W.~E.~Caswell and G.~P.~Lepage,
Phys. Lett. B \textbf{167}, 437-442 (1986)

\bibitem{Pineda:1997bj}
A.~Pineda and J.~Soto,
Nucl. Phys. B Proc. Suppl. \textbf{64}, 428-432 (1998)

\bibitem{Brambilla:2004jw}
N.~Brambilla, A.~Pineda, J.~Soto and A.~Vairo,
Rev. Mod. Phys. \textbf{77}, 1423 (2005)

\bibitem{ZinnJustin}
Jean Zinn-Justin,  
{\it Quantum Field Theory And Critical Phenomena}, Fifrth Edition,
ISBN: 9780198834625, International Series of Monographs on Physics

\bibitem{ramond2007field}
P.~Ramond, 
{\it Field Theory: A Modern Primer},
2nd Edition, 
ISBN:9780429034909,
Addison-Wesley Publishing Company, Advanced Book Program

\bibitem{Meissner:2022cbi}
U.-G.~Mei\ss{}ner and A.~Rusetsky,
{\it  Effective Field Theories},
ISBN 978-1-108-68903-8,
Cambridge University Press, 2022,

\bibitem{Feynman:1951gn}
R.~P.~Feynman,
Phys. Rev. \textbf{84}, 108-128 (1951)

\end{thebibliography}
\end{document}